\documentclass[preprint]{aastex}
\usepackage{graphicx,psfrag}
\usepackage[dvips]{color}

\newcommand{\avg}[1]{\left\langle #1\right\rangle}

\newcommand{\mint}[4]{\int_{#2}^{#3}\!\!#1\,#4}

\newcommand{\Ord}[1]{{\cal O}(#1)}

\newcommand{\pa}{\parallel}

\begin{document}

\title{Life at the front of an expanding population \\ (Research Article)}

\author{Oskar Hallatschek\altaffilmark{1}}
\affil{Max--Planck--Institute for Dynamics and Self--Organization,
   37073 G\"{o}ttingen, Germany} 
\affil{Lyman Laboratory of Physics
   and FAS Center for Systems Biology, Harvard University, Cambridge,
   Massachusetts 02138, USA }
\email{oskar.hallatschek@ds.mpg.de}

\and

\author{David R.~Nelson}
\affil{Lyman Laboratory of Physics
   and FAS Center for Systems Biology, Harvard University, Cambridge,
   Massachusetts 02138, USA }
\email{nelson@physics.harvard.edu}

\altaffiltext{1}{present address: MPI for Dynamics and
  Self--Organization, Bunsenstr. 10, 37073 G\"{o}ttingen, Germany,
  phone: +49-551-5176-670, fax: +49-551-5176-669}

\author{keywords: range expansion, stepping-stone, neutral mutation,
  genetic drift, genetic load}

\author{running head: Evolution at expanding frontiers.}

  \begin{abstract}
    The evolutionary history of many species exhibits episodes of
    habitat expansions and contractions, often caused by environmental
    changes, such as glacial cycles. These range changes affect the
    dynamics of biological evolution in multiple ways.  Recent
    microbial experiments suggest that enhanced genetic drift at the
    frontier of a two-dimensional range expansion can cause genetic
    sectoring patterns with fractal domain boundaries.  Here, we
    propose and analyze a simple model of asexual biological evolution
    at expanding frontiers to explain these neutral patterns and
    predict the effect of natural selection.  Our model attributes the
    observed gradual decrease in the number of sectors at the leading
    edge to an unbiased random walk of sector boundaries.  The long
    time sectoring pattern depends on the geometry of the frontier.
    Whereas planar fronts are ultimately dominated by only one sector,
    circular colonies permit the coexistence of multiple sectors, whose
    number is proportional, in the simplest case, to the square root of
    the radius of the initial habitat. Natural selection introduces a
    deterministic bias in the wandering of domain boundaries that
    renders beneficial mutations more likely to escape genetic drift
    and become established in a sector.  We find that the opening angle
    of those sectors and the rate at which they become established
    depend sensitively on the selective advantage of the mutants.
    Deleterious mutations, on the other hand, are not able to establish
    a sector permanently. They can, however, temporarily ``surf'' on
    the population front, and thereby reach unusually high frequencies.
    As a consequence, expanding frontiers are susceptible to
    deleterious mutations as revealed by the high fraction of mutants
    at mutation-selection balance.  Numerically, we also determine the
    condition at which the wild type is lost in favor of deleterious
    mutants (genetic meltdown) at a growing front.  Our prediction for
    this error threshold differs qualitatively from existing well-mixed
    theories, and sets tight constraints on sustainable mutation rates
    for populations that undergo frequent range expansions.
  \end{abstract}



\section{Introduction}
Population expansions in space are common events in the evolutionary
history of many species
\citep{cavallisforza93,Hewitt00,templeton2002oaa,RosenbergPWCKZF03,RamachandranDRRFC05,phillips2006iae,MathiasCurrat07142006},
ranging from biofilms to humans. Species expand from where they first
evolved, invade into favorable habitats, or move in response to
environmental changes, such as the recent climate warming, glacial
cycles, or gradients in nutrients, salinity, ambient temperature,
etc., in the case of biofilms. Some species undergo range expansions
rarely, because environments change slowly, others like epidemic
pathogens do so frequently as part of their ecology.

These range expansions cause strong differences between the genetic
diversity of the ancestral and the newly colonized regions, because
the gene pool for the new habitat is provided only by a small number
of individuals, which happen to arrive in the unexplored territory
first.  The associated alteration of the gene pool depends on the
specific demographic scenario, and encodes precious information about
the migrational history of a species.  These ``genetic footprints''
offer ways to infer, for instance, how humans moved out of
Africa~\citep{templeton2002oaa} or
species respond to climate change~\citep{Hewitt00}. However, the
underlying question, how to decipher the observed genetic patterns and
extract as much information as possible, is not yet
settled~\citep{AusterlitzJGG97,LeCorreK98,EdmondsLC04,klopfstein06,CurratE05,halla-nelson-TPB-2007}.

The most widely appreciated consequence of a range expansion is a
genetic bottleneck. Newly colonized regions are founded by a small
subset of a larger ancestral population, typically with a greater
genetic diversity. Because of this moving bottleneck at the expanding
frontier, one expects a spatial gradient in the genetic diversity
indicating the expansion direction of the ancestral population.  The
magnitude of this gradient depends sensitively on  the
population dynamics of the pioneers at the frontier (e.g., Allee
effects~\citep{Allee31}), but only weakly on the maximum population
density, also known as carrying capacity~\citep{halla-nelson-TPB-2007}.

These predicted gradients or ``clines'' in genetic diversity have
indeed been picked up by large scale genomic surveys across
populations~\citep{RefWorks:98}, and provide valuable information about
the demographic and ecological history of species.  For instance, a
frequently observed south-north gradient in genetic diversity
(``southern richness to northern purity'' \citep{RefWorks:32}) on the
northern hemisphere is thought to reflect the range expansions induced
by the most recent glacial retreat.  In the case of humans, the
genetic diversity decreases essentially linearly with increasing
geographic distance from Africa
\citep{RosenbergPWCKZF03,RamachandranDRRFC05}, which is thought to be
indicative of the human migration out of Africa.

\begin{figure}
  \psfrag{M}{1mm}
  \center
  \includegraphics[scale=0.5]{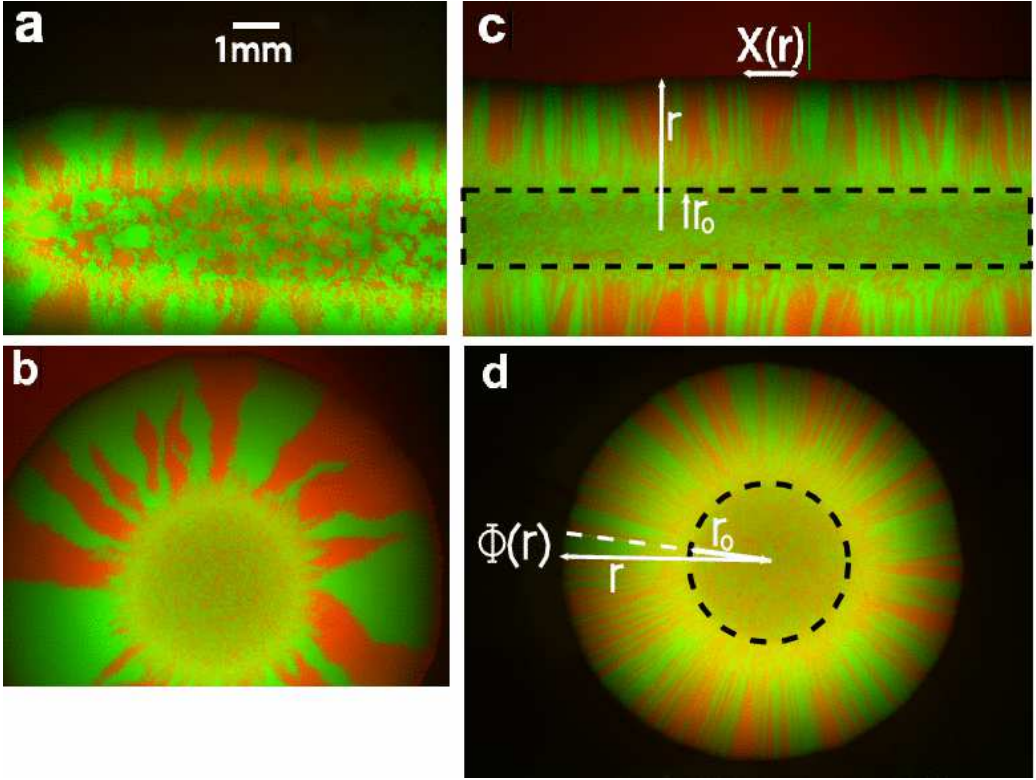}
  \caption{Competition experiments between neutral strains in growing
    bacterial ({\it E. coli}) and yeast colonies ({\it S.
      cerevisiae}).  These colonies were grown from mixtures of CFP
    (red) and YFP (green) labeled cells, which were deposited on the
    agar plates as linear (a,b) or circular (c,d) droplets (see
    Ref.~\citep{OskarHallatschek12042007} for experimental details).
    Even though both strains of each species were otherwise
    genetically identical, the growing colonies exhibit a striking
    segregation of the two neutral markers (CFP and YFP) over time.
    The dynamics of segregation is restricted to the edges of the
    colony, while, except for a gradual thickening, the interior
    distribution of CFP and YFP is essentially frozen.  Upon comparing
    the bacterial with yeast colonies (a,b versus c,d), it is apparent
    that the wandering of domain boundaries, and hence the genetic
    drift, is more pronounced for {\it E. coli} than for {\it S.
      cerevisiae}. In Figures c) and d), we have indicated quantities
    relevant to our quantitative analysis of the observed sectoring.
    The dashed box and ring in (c) and (d), respectively, indicate the
    initial size extent of the founding population. The size of
    domains at the expanding fronts are parametrized by the linear
    separation $X(r)$ as a function of the effective ``time'' $r$ in
    the linear inoculation, and by the sectoring angle $\Phi(r)$ as a
    function of radius $r$ in a circular colony.}
  \label{fig:micrographs-neutral}
\end{figure}

Until now, inference techniques have not made use of spatial
correlations in the direction \emph{transverse} to the gradient of
genetic diversity.  There is recent evidence from a microbial
experiment, however, that these spatial correlations might actually be
quite pronounced. Microbial systems have been established over the
last decade as a valuable tool to probe fundamental aspects of
evolutionary biology~\citep{elena-lenski-review-2003}. While microbial
evolution experiments were first designed for well-mixed populations,
they are now extended to spatially structured
populations~\citep{RefWorks:108}.  With these spatial systems, the
genetic impact of range expansion on the genetic diversity has
recently been measured~\citep{OskarHallatschek12042007}. It was found
that range expansion leads to a striking population differentiation
along the frontier of the advancing front. As a consequence, sectoring
patterns emerge as a distinct footprint of past range expansions, see
Fig.~\ref{fig:micrographs-neutral}.  These patterns appeared, after
two differently labeled, but otherwise identical, sub-populations were
mixed and plated on a Petri dish.  The initial liquid deposition on
the Petri dishes took the form of circular or linear droplets, the
later being inoculated off a sterile razor blade.  As these initially
well-mixed populations grew colonies, the mutant strains segregated at
the wave front and gave rise to a sectoring pattern. The growth rate
and cell mobility decrease markedly once the population wave passes
by, leaving a frozen record of gene segregation in its wake.
Qualitatively similar patterns were found in two very different
microbial species, the bacterium {\it E. coli} (a mutant strain which
lacks flagella) and brewer's yeast, {\it Saccharomyces cerevisiae} in
its haploid form.  The generic nature of the sectoring mechanism
suggests that sectors could be widespread in wild
populations~\citep{excoffier-ray-TREE08,biek07}, although direct
evidence for those patterns is limited so far~\citep{biek07}. Detecting
the trace of such spatial correlations in the genetic structure of a
species could reveal details of ancient migration patterns.

\begin{figure}
  \center
  \includegraphics[scale=0.5]{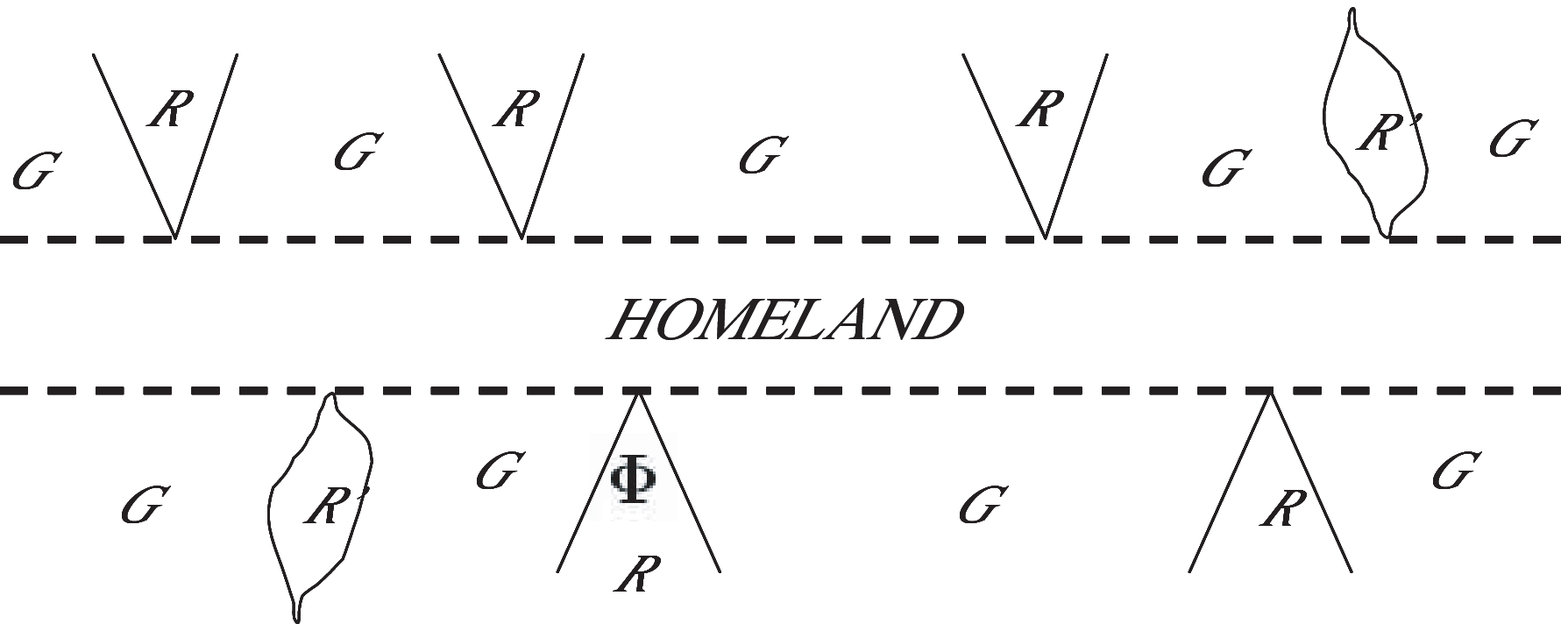}
  \caption{ Schematic illustration of the result of a linear
    inoculation along a Petri dish of a small percentage (say,
    $2-5\%$) of mutant cells with both beneficial and deleterious
    mutants (labeled red, R and R' respectively) with a majority of
    wild type (labeled green, G). (The width of the ``homeland''
    between the two dashed lines is not relevant for this discussion,
    but is determined by the amount of fluid on the razor blade and
    the capillary length of the liquid culture). As discussed in the
    text, the opening angle $\Phi$ (shown for a beneficial mutation at
    the bottom) is given by the square root of the fitness advantage.
    The potentially obscuring effects of genetic drift in the linear
    boundaries of the beneficial mutant strains are not shown. Two
    deleterious mutants that have ``surfed'' successfully for a time
    at the population front are shown as well. These surfing events
    owe their very existence to genetic drift.  }
  \label{fig:lin-inocu-scheme}
\end{figure}

Although one motivation of this paper is a better understanding of
population genetics during range expansions, another is to explore the
use of experiments such as those in
Ref.~~\citep{OskarHallatschek12042007} as an assay to measure the
effects of beneficial and deleterious mutations. Imagine mixing
together a stable ``background'' or wild type strain of bacteria or
yeast, labeled by a constitutively expressed green florescent protein
with, say, a small population (e.g., $2-5\%$ by cell number) of
mutants, all labeled red. For simplicity, assume that all favorable
mutants have an identical fitness advantage over the wild type, and
that there is a similar identical fitness detriment for the
deleterious mutants.  A possible outcome of a linear inoculation from
liquid culture on a Petri dish is indicated schematically in
Fig.~\ref{fig:lin-inocu-scheme}. As discussed in the Figure caption,
this setup allows statistics to be gathered from numerous sectoring
events in the same experiment.  Beneficial mutations give rise to
sectors with a characteristic opening angle $\Phi$, although
measurements could be obscured by genetic drift at the frontier.
Deleterious mutations, which would die out rapidly in a completely
deterministic scenario, can nevertheless take advantage of genetic
drift to ``surf'' for a period of time.  Very approximately, one can
think of the march of the two population waves away from the frontier
on the Petri dish as being like serial dilution experiments in liquid
culture, with the distance from the homeland roughly proportional to
the number of repeated ``dilutions''.  From this point of view,
advancing population fronts are a low tech, massively parallel form
of serial dilution, and could thus be used for evolution experiments.

In this article, we develop a quantitative model of biological
evolution at expanding frontiers, inspired by the microbiological
experiments described above.  Specifically, we focus on the sectoring
dynamics \emph{after} well-defined monochromatic domains have been
established in the early stages of the experiment.  Then, we argue
that allele frequencies change due to the growth and shrinkage of
sectors, which in turn reflects the competition of the deterministic
force of natural selection and random number fluctuations (genetic
drift).

\begin{figure}
  \psfrag{phi}{$\Phi$}
  \psfrag{x}{$x$}
  \psfrag{r}{$r_0$}
  \psfrag{R}{$r$}
  \psfrag{z}{$z$}
  \psfrag{M}{1mm}
  \center
  \includegraphics[scale=.5]{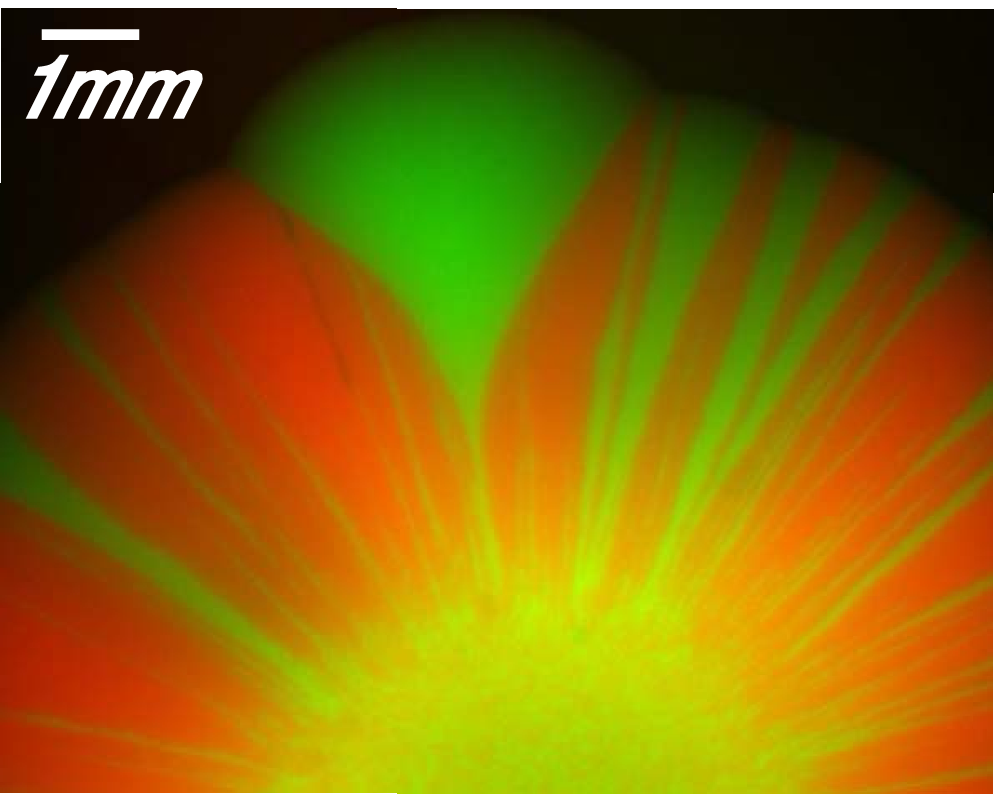}
  \includegraphics[scale=.5]{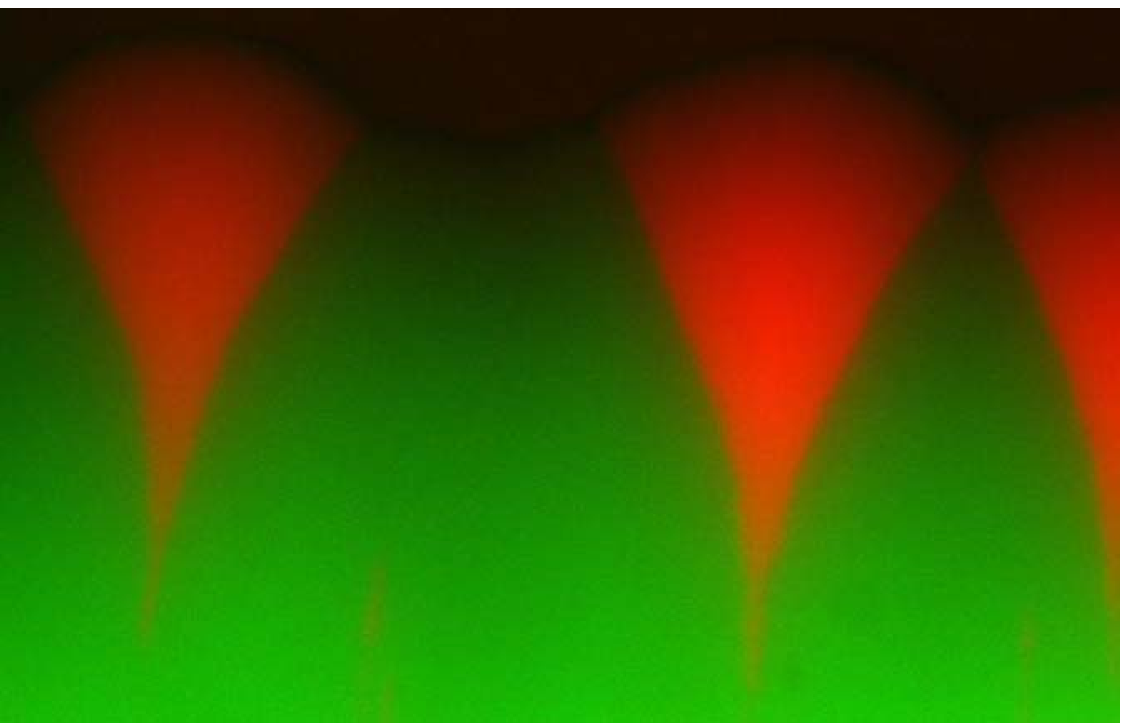} 
  \caption{ Beneficial mutations give rise to sectors with unusually
    large opening angles. a) This colony of yeast ({\it S.
      cerevisiae}) was grown from a 5:1 mixture of CFP (red) and RFP
    (green) labeled cells.  The large funnel-like green sector, which
    arose spontaneously, outgrows both wild type strains. b) A linear
    inoculation of mixture of a CFP (red) strain that has a beneficial
    mutation compared to an otherwise neutral RFP (green) strain. The
    resulting sectors of CFP mutants have similar shape and are well
    separated due to the small ratio (1:40) of mutant to wild type. }
  \label{fig:micrographs-beneficial-mutation}
\end{figure}

Even in cases where the temporal variation of sector sizes is purely
stochastic (no selection), a gradual decrease in the number of sectors
is expected as the population expands. This coarsening process occurs
because domains frequently lose contact to the wave front and no
longer participate in the colonization process
Fig.~\ref{fig:infinite-colors}.  In Sec.~\ref{sec:neutral}, we treat
this form of neutral evolution using a simple model of annihilating
random walkers, which allows us to predict the gradual decrease in the
number of sectors as the colony grows. Both linear and radial
inoculations are considered. Although our model predicts that one
allele (genetic variant) dominates for linear inoculations, the number
of surviving sectors remains finite at long times for radial
expansions, in agreement with the experiments in
Ref.~\citep{OskarHallatschek12042007}. In Sec.~\ref{sec:nat-selection},
we consider {\it biased} random walks to study the spread and ultimate
fate of beneficial mutations arising at the front of a population,
such as the one documented in
Fig.~\ref{fig:micrographs-beneficial-mutation}.  We relate the
asymptotic sector angle to fitness differences and determine the rate
at which beneficial mutations become established based on their
mutation rate. Finally, we determine the genetic load due to
deleterious mutations that accumulate in the course of a range
expansion.  Our analysis shows that, due to enhanced genetic drift,
selection is quite ineffective in purging deleterious mutations from
the invasion front. Finally, using a combination of theoretical
arguments and computer simulations, we find that a critical mutation
rate exists along a linear frontier beyond which the front population
would inevitably decline in fitness. Depending on the details of the
range expansion, our prediction for this error threshold can be much
lower than the well-known result for well-mixed ``zero-dimensional''
populations.

The emergence of well-defined domain boundaries, which underlies our
analysis, is not specific to range expansions in microbial systems, as
we argue in Sec.~\ref{sec:discussion}, which contains conclusions and
discussion.  Instead, sharp boundaries appear naturally due to the
reduction of dimensionality ($2$ to $1$) at the advancing front of a
population spreading across a surface.  Building on this hypothesis,
we discuss to what extent our results generalize beyond microbial
populations.

\begin{figure}
  \psfrag{y}{$P(x/\sigma)$}
  \psfrag{x}{$x/\sigma$}
  \center
  \includegraphics[scale=0.2]{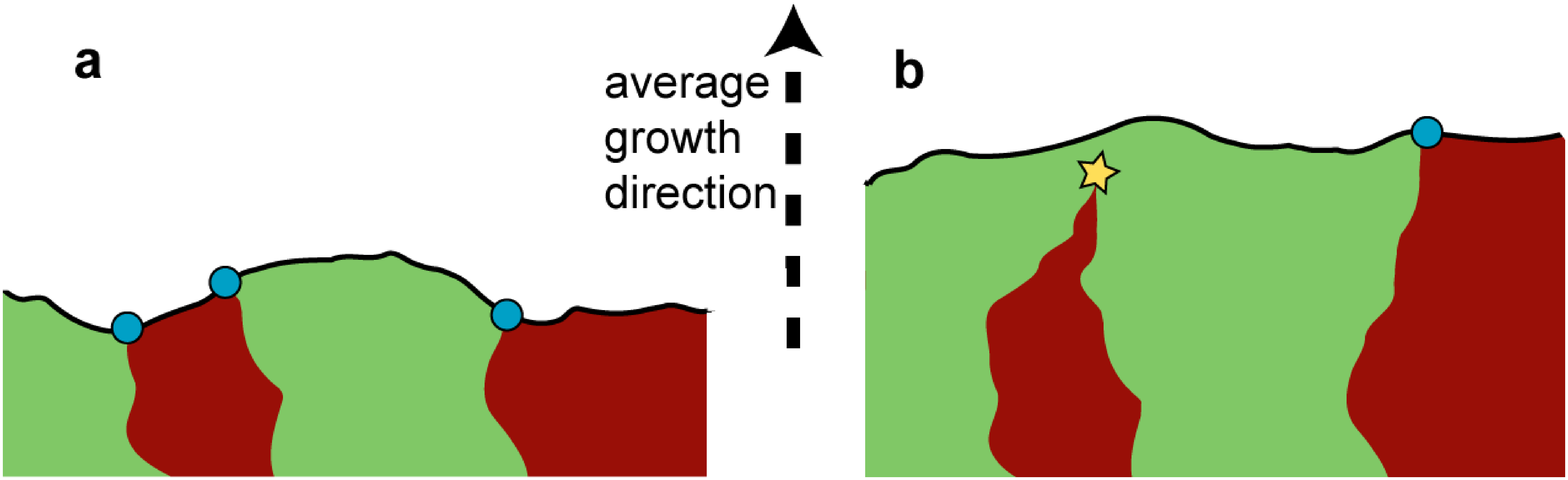}
  \caption{ An illustration of the mechanism by which the sectoring
    pattern of a microbial colony coarsens in time. (a) Four
    monochromatic domains are bounded by a moving frontier (black
    line). (b) As the colony grows further in the upward vertical
    direction, the domain boundaries follow wandering paths. By
    chance, the two domain boundaries on the left-hand side meet. As a
    consequence, the enclosed domain (red left-hand domain in (b))
    loses contact to the population front and is, henceforth, trapped
    in the bulk of the colony. We model these dynamics by replacing
    the tips of the domain boundaries by random walkers (blue circles)
    that ``live'' on the growing one dimensional edge  of the colony.
    Even though these random walkers annihilate when they meet (yellow
    star), they have a non-zero survival probability on the growing
    circumference of a circular colony.  }
  \label{fig:annihilation-sketch}
\end{figure}

\section{Materials and methods}
\label{sec:methods}
Our null model for the sectoring dynamics rests on the assumption that
the reproductive success of an individual is independent of its color.
In other words, individuals with different alleles (genetic variants)
have the same fitness.  In a strictly neutral population
mixture, domains should not grow on average, except for a linear
increase in the radial direction due to the inflating frontier in the
case of circular inoculations~\citep{OskarHallatschek12042007}.  Even
in the neutral case, however, sectors will exhibit variations in their
sizes, because of inevitable chance effects during the reproduction
process (genetic drift).  The variations in sector sizes are manifest
in the erratic path of the sector boundaries, see
Figs.~\ref{fig:micrographs-neutral}a,b. Apparently, these effects are
much more pronounced for {\it E. coli} populations than for {\it S.
  cerevisiae}. Due to these fluctuations, neighboring domain
boundaries can collide as the colony grows larger.  As a result, the
enclosed domain loses contact to the colonization frontier and is
henceforth trapped in the bulk of the colony without further
participation in the colonization process. Thus random fluctuations
due to genetic drift are responsible for the continual reduction in
sector number.

Our phenomenological model for these neutral dynamics is illustrated
for a linear inoculation in Fig.~\ref{fig:annihilation-sketch}. The
meandering ends of domain boundaries are represented by random walkers
that populate the expanding edge of the colony.  The random walkers
come in pairs, move along the advancing frontier and annihilate when
they meet.  Their trajectories in space and time describe the
meandering and coalescing of domain boundaries visible in the
microbial experiments of Fig.~\ref{fig:micrographs-neutral}.

This model of annihilating random walkers with dynamics embodied in
domain walls instead of the cells themselves serves as an
\emph{effective} description of the neutral evolutionary dynamics on
long time and length scales (see discussion). The mathematical
formulation of the model leads to generalized diffusion equations,
which could be solved analytically.

The evolutionary dynamics changes dramatically when mutations arise
close to the range margins of an expanding population that have
non-negligible fitness effects. The funnel-like sector shape on the
left side of Fig.~\ref{fig:micrographs-beneficial-mutation}, for
example, is the result of a mutation that increased the rate of
expansion in this particular green population compared to either its
unmutated green relatives or the red population. A series of similar
observations shows that beneficial mutation generically give rise to
sectors with unusually large sector angles. This phenotype implies a
bias in the diffusion process of the sector boundaries.  To account
effects of selection, we thus formulated a biased diffusion model for
the domain boundary motion. This model was solved analytically in the
limit of small mutation rates, and studied by means of spatially
explicit simulations for larger occurrence rates of deleterious
mutations.

\section{Results}
\label{sec:results}

\subsection{Neutral evolution}
\label{sec:neutral}

We first present scaling analysis for the emerging neutral coarsening
dynamics, at both planar and curved fronts. The results of a more
precise analytical calculation of the sectoring dynamics follows
thereafter.

\subsubsection{Scaling analysis}
\label{sec:scaling}
Consider the dynamics of the distance $X(r)$ between the tips of two
neighboring domain walls within a linear frontier, such as in
Fig.~\ref{fig:micrographs-neutral}a. We assume that the domain
boundary separation $X(r)$ is a continuous random
variable~\footnote{Throughout this paper, we denote random variables by
  capital letters and their values for specific realizations by
  corresponding lower case letters.} that fluctuates as if each domain
boundary carried out an independent random walk with diffusion
constant $D_X$.  This assumption implies that if the average front
position advances from $r_0$ to $r$ (see
Fig.~\ref{fig:micrographs-neutral}c) by a length increment $\Delta
r\equiv r-r_0$, then the associated increment $\Delta X=X(r)-X(r_0)$
in the domain size has zero mean and a variance $\sigma_X^2$ that
grows linearly with distance $\Delta r$,
\begin{equation}
  \label{eq:msd}
  \avg{\Delta X}=0\;, \qquad \avg{\Delta X^2}=4 D_X
  \Delta r\equiv\sigma_X^2\;. 
\end{equation}
Here, angular brackets denote an average over many realizations and
the diffusion constant $D_X$ describes the wandering of a
\emph{single} wall and has units of length$^2$/length. The extra
factor of 2 arises in Eq.~(\ref{eq:msd}) because we look at the
difference coordinate between two independent random
walks~\footnote{If the wandering of neighboring domain boundaries is
  correlated, e.g., due to interactions between the walls, we may
  consider Eq.~(\ref{eq:msd}) simply as a definition of the
  phenomenological parameter $D_X(r)$.}. Note that Eq.~(\ref{eq:msd})
also implies a random walk in real time. Indeed, if the colony expands
at a constant velocity $v$, we have $\Delta r=v \Delta t$ and may
write Eq.~(\ref{eq:msd}) equivalently as
\begin{equation}
  \label{eq:msd-rtime}
  \sigma_X^2=4 D_X v \Delta t=4 \tilde D_X \Delta t \;.  \quad \mbox{(constant
    expansion speed $v$)} 
\end{equation}
Thus, the random variable $X$ carries out a random walk in real time
$t$ with diffusion constant $\tilde D_X\equiv D_X v$. In the
following, we will employ $r$ as the effective time-like variable
rather than the real time $t$. Although the expansion velocity $v$
will vary from organism to organism, and may even be time dependent
during the early stages of a radial
expansion~\citep{murray-book-chap11-fisherwave}, $v(t)$ drops out when
the problem is formulated as in Eq.~(\ref{eq:msd}).

The random walk assumption, Eq.~(\ref{eq:msd}), forms the basic
working hypothesis of our analysis of neutral evolution. As discussed
in Ref.~\citep{OskarHallatschek12042007}, this assumption can be
violated when the interface of the expanding population is rough,
which can drive a super-diffusive wandering. This super-diffusion can
also be analyzed on the scaling level, and we shall indicate the
implied changes in Sec.~\ref{sec:discussion}. The diffusive scaling
used  here nevertheless captures the essential behavior of many
models and has the advantage of being exactly solvable. Note that our
neglect of the roughness of the interface allows us to characterize
the position of the population front by a single time-like variable
$r$ for both linear and circular inoculations (see
Fig.~\ref{fig:micrographs-neutral}).

A coalescence event occurs when two domain boundaries meet, and the
enclosed domain is displaced from the wave front in favor of the
neighboring domains.  Equivalently, we can say that the tips of domain
boundaries annihilate when they meet. The typical distance between
domain walls that evade annihilation at distance $r$ should be
comparable to the standard deviation $\sigma_X(r)$.  Hence, we show
that the average number $N(r|r_0)$ of \emph{surviving} lineages at
effective time $r$ declines with increasing $\Delta r$ as the inverse
of $\sigma_X$.  Using a pre-factor derived below in the diffusion
approximation, the number of sectors then is given by
\begin{equation}
  \label{eq:n-lin}
  N(r|r_0)= \frac{ L}{\sqrt{2\pi}\sigma_X}=
  \frac{L}{2\sqrt{2\pi D_X \Delta r}} \;,\quad \mbox{(linear inoculation)}
\end{equation}
for a neutral $50:50$ mixture of two differently labeled but neutral
sub-populations. In Eq.~(\ref{eq:n-lin}), $L$ is the total length of
the population front. Note that microscopic length scales, such as the
cell size (or, more generally, an initial domain size) do not appear
in this asymptotic formula.  The annihilation process eventually leads
to the prevalence of only one domain, or allele, after the front
advances a distance $\Delta r$ such that the distance between domain
boundaries has become of order $L$.  Although Eq.~(\ref{eq:n-lin}) is
only approximate in this regime (see below), it correctly suggests
$\Delta r\sim L^2/D_X$ as the order of magnitude of the fixation time
for a linear inoculation.

In the case of an advancing \emph{curved} population front, the
ultimate fate of the gene pool is rather different. On top of the
diffusion process, neighboring domain boundaries are subject to a
deterministic expansion caused by the changing size of the perimeter
of the population.  Thus, diffusion competes with an antagonistic
drift caused by the inflation of the perimeter, which inflates the
distance between neighboring domain walls. Although the deterministic
drift term will dominate on long times, diffusion dominates at earlier
times.

Let us consider a circular colony as the simplest example of an
advancing curved front. In this case, the perimeter grows linearly
with the radius of the colony. This radial expansion tends to increase
arc-length distances between domain boundaries.  The ends of two
neighboring domain boundaries drift apart at a speed proportional to
their distance. Thus, two neighboring random walkers are subject to
both deterministic drift and diffusion in the arc-length
parametrization of their separation.  Alternatively, we can describe
the distance between two neighboring random walkers by their
\emph{angular} distance
\begin{equation}
  \label{eq:angulardistance}
  \Delta\Phi\equiv\frac{\Delta X}{r} \;,
\end{equation}
as indicated in Fig.~\ref{fig:micrographs-neutral}d.  This change of
variables simplifies the problem to pure diffusion because sector
angles remain unchanged on average during the growth of the colony.
The angular diffusion constant $D_{\Phi}$ becomes, however, a decaying
function of the radius $r$ of the colony,
\begin{equation}
  \label{eq:drot}
  D_{\Phi}(r) \equiv \lim_{\Delta r\to0}\frac{\avg{\Delta
      \Phi^2}}{4\Delta r}= \lim_{\Delta r\to0}  \frac{\avg{\Delta
      X^2}/r^2}{4\Delta r}= \frac{D_X}{r^2} \;, 
\end{equation}
where $D_X$ is the diffusion constant that describes wall wandering
for a linear inoculation, Eq.~(\ref{eq:msd}). When the colony grows
from an initial radius $r_0$ to a final radius $r$, the mean square
angular displacement changes by
\begin{equation}
  \label{eq:msqangle}
  \sigma_{\Phi}^2\equiv\avg{\Delta\Phi^2}=4\mint{dr'}{r_0}{r}
  D_{\Phi}(r')=4 
  D_X \left(r_0^{-1}-r^{-1}\right)\;.
\end{equation}
Notice that the variance in angular distance depends on increments in
\emph{inverse} radii, as opposed to the linear dependence
Eq.~(\ref{eq:msd}) for arc-length distances. An immediate consequence
is that, when the colony grows much larger than the initial radius,
$r/r_0\gg1$, the expected mean square angular displacement stays
finite, $\sigma_\Phi\to 2\sqrt{D_X/r_0}$. At long times, the sectoring
pattern should therefore decompose into a \emph{finite} number
$N(\infty|r_0)$ given by
\begin{equation}
  \label{eq:n-approx}
  N(\infty|r_0)= \frac{L}{\sqrt{2\pi}\sigma_\Phi}=
  \sqrt{\frac{\pi r_0}{2D_X}}\;. \qquad\mbox{(circular inoculation)}  
\end{equation}
The numerical pre-factor in Eq.~(\ref{eq:n-approx}) again results from the
quantitative analysis of the sectoring statistics in the diffusion
approximation discussed below.

\subsubsection{Sector size distribution for linear and radial expansions}
\label{sec:model}
According to our model, we can view successful sectors that do not get
trapped behind the front as being bounded by random walks that evade
any collision. The size distribution of sectors should therefore be
determined by the positional distribution function of annihilating
pairs of random walkers \emph{conditional} on survival. Our goal is to
determine this distribution quantitatively.

As in the scaling discussion, we measure time by the spatial
position $r(t)$ of the frontier of the population. In the scaling
discussion, we described the size of a sector by a distance $X(r)$ in the
linear inoculation, and by an angle $\Phi(r)$ in the circular case. In
order to capture both scenarios simultaneously, let us introduce a
generalized sector size variable $Z(r)$, which is assumed to carry out a
random walk with diffusion constant $D(r)$,
\begin{equation}
  \label{eq:diffusion-approx}
  \avg{\Delta Z}=0\;,\qquad\avg{\Delta Z^2}=D(r) \Delta r \;.
\end{equation}
The specific scenarios of linear and radial inoculations can be
recovered by identifying
\begin{eqnarray}
  \label{eq:convention}
  \mbox{linear inoculation:}& Z=X\;, \qquad D(r)=D_X=\mbox{const.} \\
  \mbox{circular inoculation: }&  Z=\Phi\;, \qquad
  D(r)=D_\Phi(r)=D_X/r^2\;. 
\end{eqnarray}

The statistical properties of the sector size $Z(r)$ are described by
a diffusion equation. Let $F(z,r|z_0,r_0)$ be the probability that
$Z(r)=z$ when the front is at $r$ given that $Z(r_0)=z_0$ at the
earlier front position $r_0$. This distribution function satisfies
\begin{equation}
  \label{eq:diff-eq}
  \partial_r F(z,r|z_0,r_0)=2 D(r)\,\partial_z^2
  F(z,r|z_0,r_0)\;, 
\end{equation}
which is a direct consequence of Eq.~(\ref{eq:diffusion-approx}) and
the continuous nature of the random variable
$Z(r)$~\citep{vankampen01}. To account for the possibility of
annihilation, we impose an absorbing boundary condition at $z=0$,
\begin{equation}
  \label{eq:absorbing-bc}
  F(0,r|z_0,r_0)=0\;.
\end{equation}
The factor of two in front of the diffusion constant in
Eq.~(\ref{eq:diff-eq}) arises because $Z(r)$ is the distance between
\emph{two} random walkers.  Also note that, in order to keep the
analysis simple, we assume that space is unbounded in this section.
Finite size effects can lead to the important effect of fixation as
discussed in Sec.~\ref{sec:fixation}.

\begin{figure}
  \psfrag{t}{$r_0$}
  \psfrag{tau}{$r$}
  \center
  \includegraphics[scale=0.8]{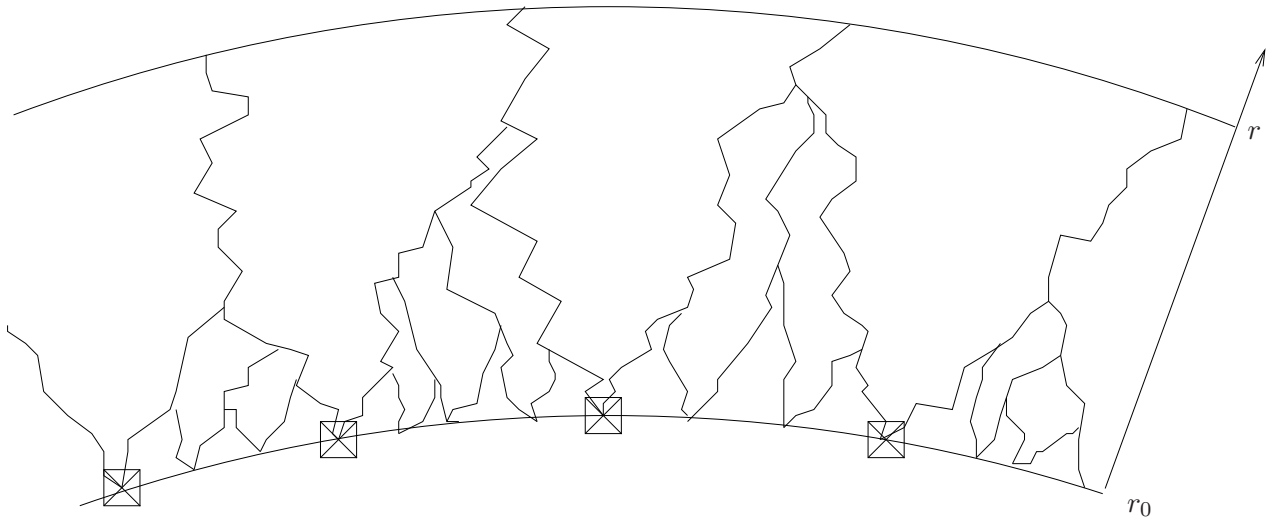}
  \caption{ A sketch of the sectoring pattern emerging from an
    (hypothetical) initial population in which each individual is
    labeled differently. The curved colony grows from initial radius
    $r_0$ to radius $r$. Each surviving lineage emerges from a single
    founder cell denoted by a cross. 
    \label{fig:infinite-colors}}
\end{figure}

The absorbing boundary condition Eq.~(\ref{eq:absorbing-bc}) can be
fulfilled exactly by writing the solution as
\begin{equation}
  \label{eq:crw-solution}
  F(z,r|z_0,r_0)=G(z,r|z_0,r_0)-
  G(-z,r|z_0,r_0) \;,  
\end{equation}
in terms of the solution $G(z,r|z_0,r_0)$ of the diffusion equation
(\ref{eq:diff-eq}) \emph{without} annihilation. Diffusion without
annihilation, on the other hand, is well-known to be described by a
Gaussian probability distribution,
\begin{equation}
  \label{eq:gaussian-distribution}
  g(\Delta z,\Delta r)\equiv G(z_0+\Delta z,r_0+\Delta
  r|z_0,r_0) =\frac{\exp\left(\frac{-\Delta
        z^2}{2\sigma^2}\right)}{\sqrt{2\pi  
      \sigma^2}} \;,
\end{equation}
where $\sigma$ is the standard deviation of the random variable $Z$
accumulated from frontier distance $r_0$ to distance $r$,
\begin{equation}
  \label{eq:stddev-general}
  \sigma^2(r,r_0)\equiv4\mint{dr'}{r_0}{r}D(r')=\left\{ 4D_X(r-r_0)
    \mbox{ linear} \atop 4D_X(r_0^{-1}-r^{-1})\mbox{ radial} \;. \right.
\end{equation}
This quantity was evaluated in the scaling analysis in
Eqs.~(\ref{eq:msd}) and (\ref{eq:msqangle}) for the linear and circular
inoculation, respectively. Upon combining Eqs.~(\ref{eq:crw-solution})
and (\ref{eq:gaussian-distribution}), we obtain
\begin{eqnarray}
  \label{eq:scalingsolution}
  F(z_0+\Delta z,r_0+\Delta
  r|z_0,r_0)&=&g(\Delta z,\Delta r) -g(-2z_0-\Delta z,\Delta r) \nonumber \\
  &=&\frac{\exp\left(-\frac{\Delta z^2}{2\sigma^2}\right)}{\sqrt{2\pi
      \sigma^2}}\left[ 1- \exp \left(\frac{-2z_0(z_0+\Delta z)}{
      \sigma^2}\right)\right]   \;.
\end{eqnarray}
This result can be used to predict the sector size distribution that
emerges when a linear or circular colony grows from a finely divided
mixture of differently labeled sub-populations. Suppose that the front
of the colony advances from position $r_0$ to position $r>r_0$.  Then,
every point along the edge of the population sends out domain
boundaries in a tree-like web that gradually coalesces, as illustrated
in Fig.~\ref{fig:infinite-colors}.  Due to this coarsening process,
the number of sectors $N(r|r_0)$ at the front is a decreasing function
of time $r$, whose ensemble average we seek to determine.  Consider
first an ``infinite alleles model''~\citep{RefWorks:33} where each cell
of the founder population is labeled differently. In this case, each
surviving sector at $r$ originates at earlier effective time $r_0$ in
a single individual, the most recent common ancestor (denoted by
a cross in Fig.~\ref{fig:infinite-colors}).

\begin{figure}
  \psfrag{y}{$\sigma P(z/\sigma)$}
  \psfrag{x}{$z/\sigma$}
  \psfrag{a}{\hspace{-.3cm}$L/\sigma=1.5$}
  \psfrag{b}{$2.0$}
  \psfrag{c}{\hspace{-.2cm}$2.5$}
  \psfrag{d}{\hspace{-.23cm}$3.0$}
  \psfrag{e}{\hspace{-.25cm}$3.5$}
  \psfrag{f}{\hspace{-.2cm}$4.0$}
  \center
  \includegraphics[scale=0.8]{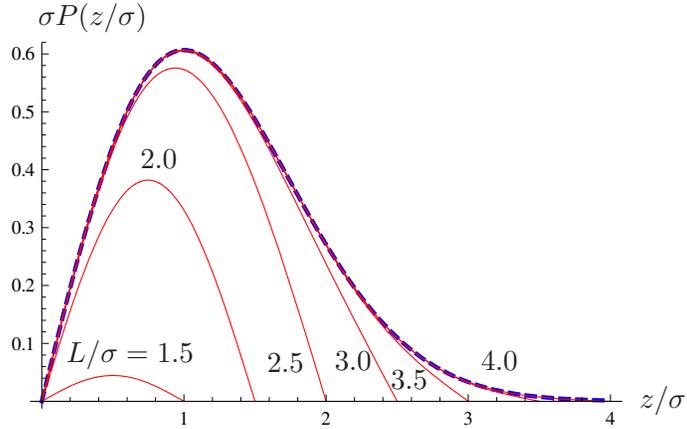}
  \caption{ Sector size distribution predicted by the
    annihilating random walker model. The horizontal axis is the ratio
    of sector-size $z$ and the positional standard deviation
    $\sigma(r,r_0)$ accumulated by a random walk between times $r_0$
    and $r$, as defined in Eq.~(\ref{eq:stddev-general}).  Full lines
    represent exact solutions with the parameter $L/\sigma$ increasing
    in steps of $0.5$ in order of increasing peak height, as obtained
    from Eq.~(\ref{eq:phi-exact}). For $L/\sigma\geq 4$, the exact
    solution is hardly distinguishable from the asymptotic result
    Eq.~(\ref{eq:sector-size-distribution}) (dashed line), which is
    independent of $L/\sigma$.  The area under these curves represents
    the probability that a randomly chosen sector has not (yet) reached
    fixation.  For small $L/\sigma \ll1$, the area approaches $0$
    because the fixation probability of a sector approaches $1$.}
  \label{fig:sprob}
\end{figure}

Consequently, a sector is generated by two domain boundaries that
start from the same location of the common ancestor at initial time
$r_0$.  Hence, the sector size distribution is simply given by the
probability distribution function of random walkers that evade
annihilation given that they start (almost) at the same place. This
distribution can be obtained by normalizing the small $z_0$ expansion
of $F(z,r|z_0,r_0)$ in Eq.~(\ref{eq:scalingsolution}). We thus obtain
the sector size distribution
\begin{equation}
  \label{eq:sector-size-distribution}
  P(z, r|r_0)= \frac{ z}{\sigma^2} \exp\left( -\frac{
      z^2}{2 \sigma^2}  \right)\;,\qquad z\geq0\;.
\end{equation}
$P(z,r|r_0)$ is the probability that a randomly chosen sector sampled
at time $r$ has size $z$ in the infinite alleles model. The product
$\sigma P$ is plotted as a function of the dimensionless variable
$z/\sigma$ in Fig.~\ref{fig:sprob} (dashed line). Note that
Eq.~(\ref{eq:sector-size-distribution}) assumes that the population
front is unbounded. For a circular inoculation, $0\leq z \leq 2\pi$,
Eq.~(\ref{eq:sector-size-distribution}) is only valid in the limit
$\sigma(r,r_0)\ll1$. The more complicated exact distribution for this
bounded case, Eq.~(\ref{eq:phi-exact}), is derived in
Sec.~\ref{sec:fixation} below.

The mean of the distribution Eq.~(\ref{eq:sector-size-distribution})
is given by
\begin{eqnarray}
  \label{eq:sector-size}
  \avg{Z(r|r_0)}&=&\sqrt{\frac\pi 2}\sigma(r,r_0) \\
  &=&\left\{\sqrt{2 \pi D_X (r-r_0)}\qquad \mbox{(linear inoculation)}
      \atop \sqrt{2  \pi D_X \left(r_0^{-1}-r^{-1}\right)}
      \qquad \mbox{(circular inoculation)} \;, \right.  \nonumber 
\end{eqnarray}
and represents the average size of a sector. Equivalently, it is the
average distance between two non-colliding random walkers that
initially start out from (almost) the same place.  The numerical
pre-factor may be compared with the expected size $\sigma\sqrt{2/\pi}$
of a sector in the \emph{absence} of annihilation, as can easily be
shown from Eq.~(\ref{eq:gaussian-distribution}). Thus, the average
separation of surviving annihilating random walks is a factor of
$\pi/2\approx 1.57$ larger than of ``phantom'' random walks that can
pass freely through each other.  Intuitively, this factor represents
an effective repulsion between two random walks that must avoid
collision to survive.

The average number $N(r|r_0)$ of sectors measured in many repeated
experiments of a given kind (circular or linear), is given by the
ratio of total size of the front $L$ and average sector size,
\begin{equation}
  \label{eq:nr-of-sectors}
  N(r|r_0)=\frac{L}{\avg{Z(r|r_0)}}=
  \sqrt{\frac{2}{\pi}}\frac{L}{\sigma} \qquad\mbox{(infinite
    alleles model)}\;.
\end{equation}
Here, $L$ would be the total length of the population front for a
linear inoculation. In the circular case, however, one would choose
$L=2\pi$. The standard deviation then approaches
$\sigma_\Phi=2\sqrt{D_X/r_0}$ for $r\to \infty$, so that
$N(\infty|r_0)=\sqrt{2\pi D_X/r_0}$.~\footnote{Provided that sector
  interactions (neglected here) are short range, we expect that
  corrections to Eq.~(\ref{eq:nr-of-sectors}) are of order $\sqrt{N (r | r_0
  )}$.}

These results, however, only hold in the infinite alleles model. When
we relax the assumption that every founder individual has a different
color, the average number of sectors will be less. For instance, if
the initial population is labeled by only two colors in equal
proportions, as in the experiment in
Ref.~\citep{OskarHallatschek12042007}, two neighboring sectors of the
infinite alleles model will have the same color with probability
$1/2$. Thus, there will be half as many sector boundaries as in the
infinite alleles model. Accordingly, the average number of sectors in
the two alleles model will be given by
$N(r|r_0)=L(r)/\avg{2Z(r|r_0)}$, which gives the pre-factors used in
Eqs.~(\ref{eq:n-lin}) and (\ref{eq:n-approx}) of our scaling analysis.
More generally, we may consider an initially well-mixed population, in
which two randomly chosen individuals have a different color with
probability $H$, known as heterozygosity~\citep{RefWorks:33}.  Then the
number of sectors will be given by the result for the infinite alleles
model, Eq.~(\ref{eq:nr-of-sectors}), multiplied by $H$.

\begin{figure}
  \psfrag{y}{$f(\sigma/L)$} \psfrag{x}{$\sigma/L$} \center
  \includegraphics[scale=0.8]{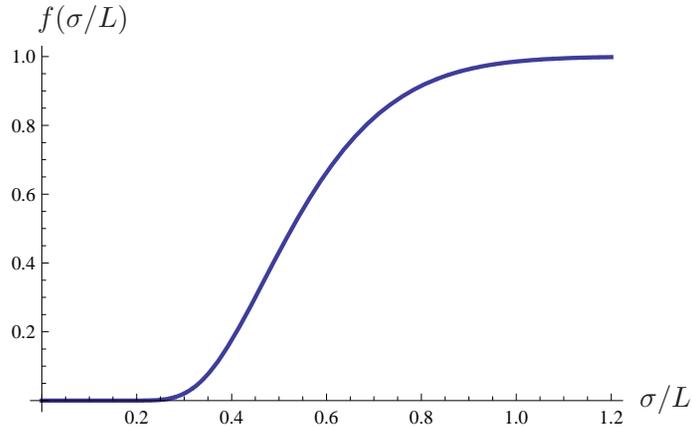}
  \caption{ The graph describes how the probability of fixation depends
    on the amount of domain wall wandering. The function $f(\sigma/L)$
    represents the probability that fixation is reached at a front of
    size $L$ given that the accumulated variance of a single domain
    boundary is $\sigma^2(r,r_0)$, as defined in
    Eq.~(\ref{eq:stddev-general}).  }
  \label{fig:fprob}
\end{figure}

\subsubsection{Finite front size - probability of fixation}
\label{sec:fixation}
Our previous results were derived under the assumption that the
frontier is very large, or rather, that sectors are too small to
``notice'' that the front size is actually limited.  However, when a
sector grows up to a size comparable to the dimension $L$ of the
front, we have to account for the possibility that the sector could
take over the entire front and reach fixation. In this section, we
determine the probability of fixation assuming periodic boundary
conditions, as would be appropriate for a radial inoculation, or
for microorganisms growing from a linear inoculation around the waist
of a cylinder.

\begin{figure}
  \center
  \includegraphics[scale=0.5]{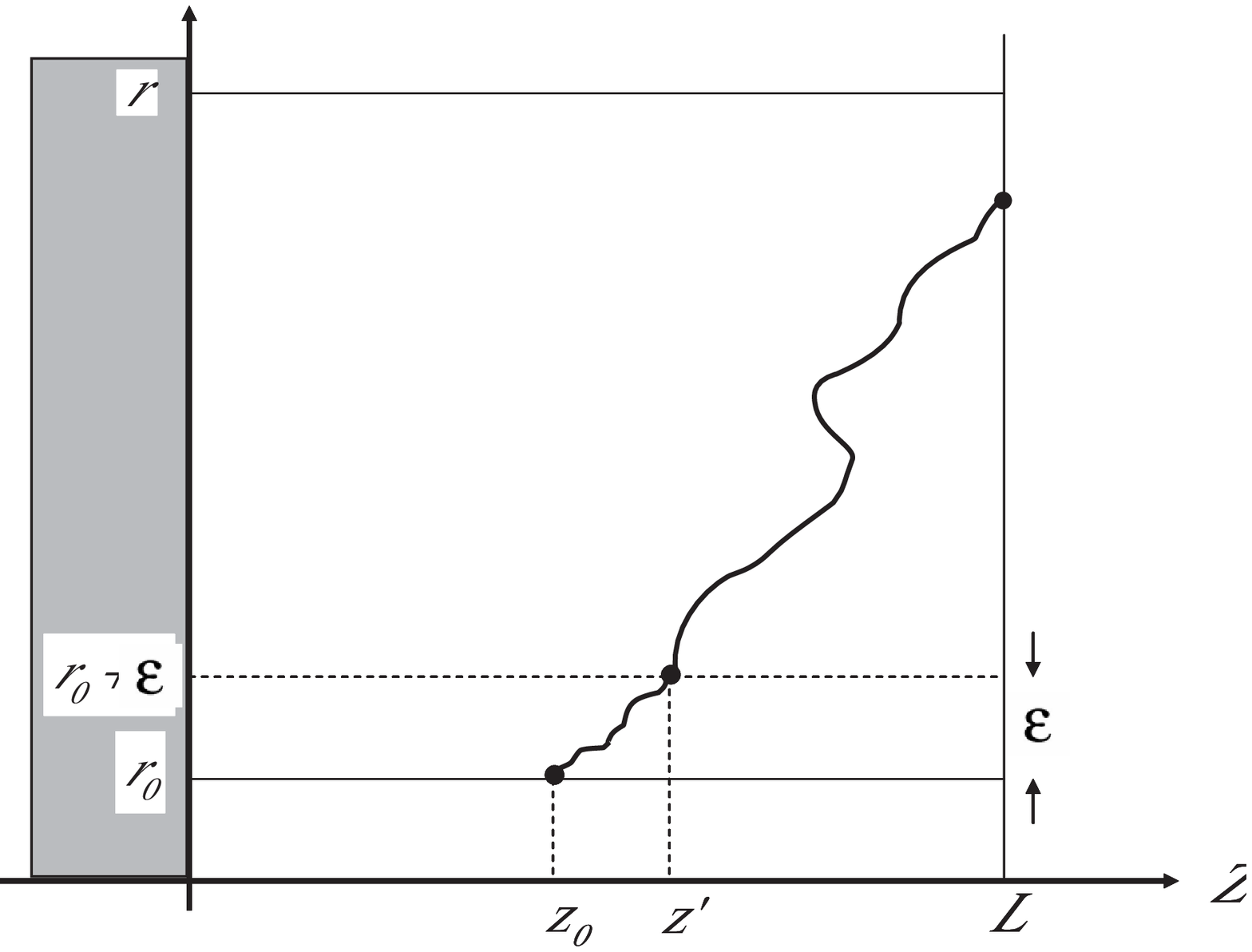}
  \caption{  Graphical representation of Eq.~(\ref{eq:chap-kol}) for
    the probability $u(r|z_0,r_0)$ a sector fixes at any frontier
    position up to and including $r$, given that the width $Z$ of the
    sector was $z_0$ at earlier frontier position $r_0$.  Fixation
    occurs when the fluctuating width (black line) of a sector matches
    the system size $L$.}
  \label{fig:wandering-sector-size}
\end{figure}

As a first step towards this goal, let us determine in the infinite
alleles model the probability $u(r|z_0,r_0)$ that a sector has swept
to fixation, $Z=L$, before the frontier has reached position $r$, given
that the sector was a size $Z(r_0)=z_0$ at initial ``time'' $r_0$.
Here, $L=\pi d$, for a linear cylinder of diameter $d$, and $L=2\pi$
for a radial inoculation. To this end, it is convenient to derive from
the diffusion equation Eq.~(\ref{eq:diff-eq}) an equivalent
differential equation for $u(r|z_0,r_0)$. This can be done, in analogy
with derivations of the Kolmogorov backward equation in population
genetics~\citep{RefWorks:33}, by writing
\begin{equation}
  \label{eq:chap-kol}
  u(r|z_0,r_0)=\mint{dz'}{}{}
  u(r|z',r_0+\epsilon)F(z',r_0+\epsilon|z_0,r_0) \;,
\end{equation}
which follows from the Markov property of the problem, see
Fig.~\ref{fig:wandering-sector-size}. Upon expanding the right hand
side in $\epsilon$ and using Eq.~(\ref{eq:diff-eq}), we obtain
\begin{eqnarray}
  \label{chap-kol2}
  u(r|z_0,r_0)&=&\int{}{}{dz'} \left[u(r|z',r_0)+\epsilon \partial_{r_0}
    u(r|z',r_0)\right]\\
  & &\times\left[ F(z',r_0|z_0,r_0)+\epsilon 2D(r_0) \partial_{z'}^2
   F(z',r_0|z_0,r_0)+\mbox{O}(\epsilon^2) \right] \nonumber  
\end{eqnarray}
After noting that $F(z,r_0|z_0,r_0)=\delta(z-z_0)$ and integrating by
parts, we find
\begin{eqnarray}
  \label{eq:pde-surv}
  \partial_{r_0} u(r|z_0,r_0)&=&- 2 D(r_0)\partial_{z_0}^2
  u(r|z_0,r_0)  
\end{eqnarray}
to order $\Ord{\epsilon}$. Note that all derivatives in
Eq.~(\ref{eq:pde-surv}) act on the coordinates $z_0,r_0$
characterizing the initial conditions.

We seek a solution to Eq.~(\ref{eq:pde-surv}) subject to the ``final''
condition $u(r|z_0,r)=0$ (no fixation if the frontier does not advance),
and two boundary conditions,
\begin{equation}
  \label{eq:annihi-bc}
  u(r|0,r_0)=0 \qquad u(r|L,r_0)=1\;.
\end{equation}
The first boundary condition accounts for the annihilation of a sector
as $z_0\to 0$. The second guarantees that the fixation probability is
1 for all $r>r_0$ if the sector already spans the entire front  at
initial  position $r=r_0$.

Note that the function $z_0/L$ satisfies both Eq.~(\ref{eq:pde-surv})
and the boundary conditions Eq.~(\ref{eq:annihi-bc}), which motivates
the following ansatz,
\begin{equation}
  \label{eq:ansatz}
  \Delta u(r|z_0,r_0) \equiv 
  u(r|z_0,r_0)-z_0/L=\sum_{n=1}^{\infty}a_n(r_0)W_n(z_0)\;,
\end{equation}
with coefficients
\begin{equation}
  \label{eq:coefficients}
  a_n(r_0)=\mint{dz_0}{0}{ L} W_n(z_0) \Delta u(r|z_0,r_0)\;,
\end{equation}
given in terms of a complete orthonormal basis set of sine functions
\begin{equation}
  \label{eq:sine-modes}
  W_n(z)=\sqrt{\frac{2}{L}}\sin(q_n z) \qquad q_n\equiv\frac {n\pi}{L} \;.
\end{equation}
The function $\Delta u(r|z_0,r_0)$ represents the difference between
the time-dependent solution, $u(r|z_0,r_0)$, and the linear steady
state solution, $z_0/L$.  The expansion of $\Delta u(r|z_0,r_0)$ in
terms of sine-modes guarantees that the boundary conditions
Eq.~(\ref{eq:annihi-bc}) are satisfied.

Inserting the ansatz Eq.~(\ref{eq:ansatz}) into
Eq.~(\ref{eq:pde-surv}) gives the evolution of the mode
amplitudes $a_n(r_0)$ with $r_0$,
\begin{equation}
  \label{eq:EOM-modeamplitudes}
  \partial_{r_0} a_n(r_0)=2 q_n^2 D(r_0)a_n(r_0) \;.
\end{equation}
Upon integrating from $r$ to $r_0$, we obtain
\begin{equation}
  \label{eq:time-evol-mode-amplitudes}
  a_n(r_0)=a_n(r) e^{- q_n^2\sigma^2/2}\;,
\end{equation}
with $\sigma=\sigma(r,r_0)$ being the standard deviation defined in
Eq.~(\ref{eq:stddev-general}). The pre-factor $a_n(r)$ is determined
by imposing the final condition $u(r|z_0,r)=0$ on
Eq.~(\ref{eq:coefficients}),
\begin{eqnarray}
  \label{eq:final-cond-fixaiton}
  a_n(r)&=&\mint{dz_0}{0}{L}W_n(z_0)\;\Delta  u(r|z_0,r) \\
  &=&-\sqrt{\frac{2}{L}}\mint{dz_0}{0}{L}\sin(q_n z_0)\frac {z_0}L \\ 
  &=&\sqrt{\frac{2}{L}}\left\{{q_n^{-1}\;,\qquad n \mbox{ even}
      \atop -q_n^{-1}\;, \qquad      \mbox{otherwise}} \right. \;.
\end{eqnarray}
Hence, the solution for the fixation probability of a sector of
initial size $z_0$ reads
\begin{equation}
  \label{eq:surv-proba}
  u(r|z_0,r_0)=\frac {z_0}L+\sum_{n=1}^{\infty}\frac{2(-1)^n}{q_n L}\sin\left(
    q_n z_0  \right)e^{- q_n^2\sigma^2/2}\;.
\end{equation}
Clearly, the absolute fixation probability goes to zero as the initial
size $z_0$ of the sector decreases, because small sectors almost
always get trapped by the absorbing boundary condition on the left
side of Fig.~\ref{fig:wandering-sector-size}. 

In the infinite alleles model, we have an initially very large number
$L/z_0$ of very tiny sectors, $z_0\to 0$, that all compete for taking
over the colonization front. One and only one sector can reach
fixation. In order to study the time dependent fixation probability of
this successful sector we condition on ultimate survival. We ask, what
is the probability that fixation is complete by the ``time'' $r$
provided that the conisdered sector reaches fixation? In the infinite
alleles model, this probability follows from Eq.~(\ref{eq:surv-proba})
after a normalization,
\begin{eqnarray}
  \label{eq:fixproba-as-fct-of-sigma}
  f(r|r_0)&\equiv&\lim_{z_0\to 0}\frac{u(r|z_0,r_0)}{u(\sigma\to\infty|z_0,r_0)} \\
  &=&1+2\sum_{n=1}^{\infty}(-1)^n e^{-q_n^2 \sigma^2/2} \nonumber \\
  &=&\vartheta_4\left[0,\exp\left(\sigma^2 \pi^2/(2
        L^2)\right)\right] \;,
  \label{eq:fixproba-as-fct-of-sigma-3} 
\end{eqnarray}
where $\vartheta_n[z,q]$ are the Jacobi theta functions.  The function
$f(r|r_0)$ represents the probability that a colonization experiment
running from time $r_0$ until time $r$ reaches fixation in the
infinite alleles model.  The denominator in
Eq.~(\ref{eq:fixproba-as-fct-of-sigma}) is a normalizing factor that
ensures $f(r|r_0)\to 1$ as $\sigma(r)\to\infty$. The dependence of
$f(r|r_0)$ on $\sigma/L$ is shown in Fig.~\ref{fig:fprob}.

At linear fronts, the accumulated standard deviation
$\sigma=\sigma_X\propto \sqrt{\Delta r}$ increases without bound, so
that fixation is inevitable at long times. For this case, (again
restricting our attention to the infinite alleles model) we can
determine the average ``time'' or frontier distance to fixation,
$\avg{\Delta r_{fix}}$, by an integral over $f(r|r_0)$,
\begin{eqnarray}
  \avg{\Delta r_{fix}}&=&\mint{dr}{r_0}{\infty}(r-r_0) \partial_r f(r|r_0) \\
  &=&\mint{dr}{r_0}{\infty}\left[1-f(r|r_0)\right] \\
  &=&\frac{L^2}{12 D_X} \;,\qquad \mbox{(linear
    inoculations)}   \label{eq:time-to-fix}
\end{eqnarray}
where we used Eq.~(\ref{eq:msd}) for the variance $\sigma_X^2$ and
Eq.~(\ref{eq:fixproba-as-fct-of-sigma-3}) to evaluate the integral in
the second line, and the boundary condition $\lim_{r\to
  \infty}f(r|r_0)\to 1$.  Note that the pre-factor in
Eq.~(\ref{eq:time-to-fix}) holds for periodic boundary conditions
only, but can be evaluated for other boundary conditions along the
same lines, see e.g.~Ref.\citep{Redner-book2007}. Our derivation
assumes the infinite alleles scenario, and we have not yet found an
exact result for a finite number of colors. We expect, however, that
the fixation time is linearly dependent on the heterozygosity $H$ of
the initial population, at least for small heterozygosities.  Our
approximate argument is as follows.  Suppose the initial population is
an unbalanced binary mixture with $H\ll1$.  Then, with high
probability $(1-H)$ the majority population will take over in a short
time of order $(H L)^2/D$. On the other hand, with a low probability
$H$ the colony will be taken over by the minority population, and this
will take a time comparable to the fixation time $L^2/D$ in the
infinite alleles model.  As a consequence, these rare events of the
fixation of the minority population dominate the avarage fixation
time, which should therefore be of order $H L^2/D$.

Finally, we also generalize the sector size distribution
Eq.~(\ref{eq:sector-size-distribution}) to finite-sized frontiers. To
this end, we solve Eq.~(\ref{eq:diff-eq}) under an additional
absorbing boundary condition,
\begin{equation}
  \label{eq:boundary-condition-L}
  \left.  F(z,r|z_0,r_0)\right|_{z=L}=0
\end{equation}
which amounts to disregarding a sector once it reaches fixation. In
Supplementary Text 1, we show that the solution to this problem
can again be found in terms of the sine modes introduced in
Eq.~(\ref{eq:sine-modes}), specifically
\begin{equation}
  \label{eq:phi-exact}
  F(z,r|z_0,r_0)=\frac 2L \sum_{n=1}^{\infty}\sin(q_n x)\sin(q_n
  z_0)  e^{-q_n^2\sigma^2/2} \;.
\end{equation}
As in the unbounded case, the sector size distribution is found by
letting $z_0\to0$,
\begin{equation}
  \label{eq:phi-exact-2}
  P(z,r|r_0)\propto \sum_{n=1}^{\infty}q_n L \sin(q_n x)
  e^{-q_n^2\sigma^2/2} \;. 
\end{equation}
In Fig.~\ref{fig:sprob}, we plot this sector size distribution
normalized by the probability that a sector has not (yet) reached
fixation.  The results are indistinguishable from the asymptotic
result Eq.~(\ref{eq:sector-size-distribution}) for $L/\sigma\leq 4$.

\subsection{Natural selection}
\label{sec:nat-selection}
\begin{figure}
  \psfrag{front}{frontier position}
  \psfrag{T}{time} 
  \psfrag{M}{$v_\perp$}
  \psfrag{theta}{\hspace{.2cm}$\Phi$}
  \psfrag{vstar}{$\propto v^\star \cos(\Phi/2)$}
  \psfrag{regular}{$\propto v$}
  \psfrag{tau}{$r$}
  \psfrag{t}{$r_0$}
  \center
  \includegraphics[scale=0.3]{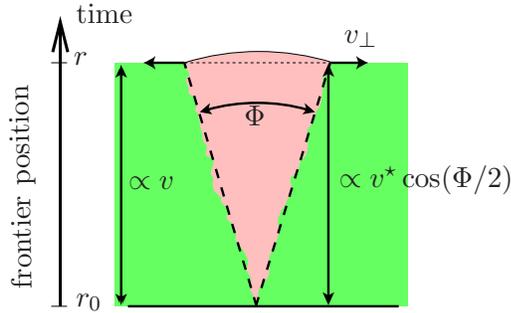}
  \caption{ A simple model of a beneficial sector within a planar wave
    front in the long-time limit. We assume that a beneficial mutation
    arises at front position $r_0$ and is able to overcome the
    short-time genetic drift at the front causing wobble in the domain
    boundaries. At long times the beneficial sub-population segregates
    and forms its own sector growing at radial velocity $v^\star$, which is
    assumed to be larger than wild-type growth speed $v$.  Requiring
    that the kinks at the interface, where the wild type and mutant
    population meet, are reached by both populations at equal times,
    leads to $\cos(\Phi/2)=v/v^\star$.}
  \label{fig:beneficial-mutation}
\end{figure}

The above model can be extended to describe the effect of natural
selection on mutations occurring during a range expansion, such as the
one documented in Fig.~\ref{fig:micrographs-beneficial-mutation}. For
simplicity, we remove complications associated with inflation by
focusing on linear inoculations. In the previous section on neutral
evolution, we ensured that each sector had the same chance of survival
by requiring that the expected change in sector size vanishes,
$\avg{\Delta X}=0$, see Eq.~(\ref{eq:msd}).  This neutrality
assumption is no longer valid if a sector is generated by mutants that
have a different fitness than the wild type.  To describe the fate of
those non-neutral sectors, it is natural to assume that a selective
advantage biases the sector growth. A sector harboring beneficial
mutations will tend to increase its size, whereas a deleterious
mutation will decrease the size. Upon setting $\Delta r=r-r_0$ as
before, we thus generalize Eq.~(\ref{eq:msd}) to
\begin{equation}
  \label{eq:msd-selection}
  \avg{\Delta X}=2 m_\perp \Delta r\;, \qquad\avg{\Delta X^2}=4 D_X
  \Delta r\;. 
\end{equation}
Here, we have assumed that selection does not alter the diffusion
constant $D_X$, which seems reasonable, at least in the case of weak
selection. The new parameter $2 m_\perp$ describes the increase in the
mean sector size due to selection and has units of a slope,
length/length.  The factor of $2$ indicates that each of the two
boundaries exhibits an average lateral drift of $m_\perp$, adding up to
a total sector growth rate of $2m_\perp$. In terms of measurable units, the
bias parameter $m_\perp$ is given by the ratio of the velocity
$v_\perp$ of a sector growth at right angles to the direction in which
the front is advancing and the velocity $v$ of the wild type range expansion,
\begin{equation}
  \label{eq:nu_perp}
  m_\perp\equiv \frac{v_\perp}{v}=\tan(\Phi/2)\;.
\end{equation}   
The second equality describes the relation between $m_\perp$ and the
opening angle $\Phi$ of the sector at long times, which can be
perceived from the sketch in Fig.~\ref{fig:beneficial-mutation}. For
weak selection, $\Phi$ will be small, and we can think of $m_\perp$ as
being just half the (asymptotic) opening angle of the sector. This
opening angle depends on the relation between the fitness effect of
the mutation and the demographic expansion process.

Although the relation between $\Phi$ (or $m_\perp$) and selective
advantage is complicated in general, it can be determined in
two simple cases, both of which should be realizable for populations
of microorganisms on a Petri dish, and for other range expansions as
well.  We assume that the only phenotypic effect of a beneficial
mutation is an increase in the expansion velocity
$\nu\to\nu^\star=f(s) \nu$, where $f(s)>1$.  Here, $s$ is the
usual selective advantage, defined as the ratio of growth rates $a^\star$ and
$a$ of a population of organisms \emph{at the frontier},
$a^\star/a=1+s$.  In principle, $a$ (or $a^\star$) could be measured
directly by monitoring cell divisions under a microscope, as was done
in the sectoring experiments of Ref.~\citep{OskarHallatschek12042007}.
However, the relation between the more easily measured front growth
velocities and $s$ is known when population number fluctuations (i.e.,
effects of genetic drift) at the front are weak.  One then expects the
range expansions to be described by Fisher population
waves~\citep{murray-book-chap11-fisherwave}, for which $v=2\sqrt{D a}$
and $v^\star=2\sqrt{D^\star a^\star}$, where $D$ and $D^\star$ are
population diffusion constants at the frontier.  If the beneficial
mutation does not change the diffusion constant, we then have
\begin{equation}
  \label{eq:f-deterministic}
  f(s)=\sqrt{1+s}\;.
\end{equation}
Wakita et al.~have in fact measured a square root dependence of the
growth velocity on the nutrient concentration in plates of bacillus
subtilis~\citep{wakita-bac-subt}, consistent with
Eq.~(\ref{eq:f-deterministic}) above when the growth rate is
proportional to the nutrient concentration.  Results are also known in
the limit when genetic drift dominates growth at the
frontier~\citep{RefWorks:30}. In this limit, the growth velocities are
proportional to doubling rates at the frontier.  Assuming all other
quantities are the same for the mutant and wild type, we then have
\begin{equation}
  \label{eq:f-deterministic-b}
  f(s)=1+s\;.
\end{equation}
See Supplementary Text 2 for a more detailed discussion.

More generally, we expect that $f(s=0)=1$, and a Taylor series
expansion about of the form
\begin{equation}
  \label{eq:taylor}
  f(s)=1+cs+\dots
\end{equation}
where $c>0$ and $s$ are small. Under the conditions described above,
we have $c=1/2$ and $c=1$ for weak and strong genetic drift
respectively.  

Now suppose that a mutation arises (or a mutant cell is inserted by
hand) close to the linear front of a wild type population in a Petri
dish, and is able to overcome the critical initial phase, where
selection is weak compared to genetic drift. For long times, the
mutant sub-population will then form a sector that grows by the factor
$f(s)$ faster into the unoccupied space than the wild type, as
discussed above. Assume that, after initial transients, a
time-independent sector angle forms.  Again appealing to
Fig.~\ref{fig:beneficial-mutation}, we see that the mutant sector
asymptotically approaches an opening angle $\Phi$ that satisfies
\begin{equation}
  \label{eq:theta}
  \cos(\Phi/2)=\frac{v}{v^\star}=\frac{1}{f(s)}\approx 1- c s \;.
\end{equation} 
Thus, for this ``geometric'' model of the opening angle $\Phi$, the
drift parameter $m_\perp=v_\perp/v$ in Eq.~(\ref{eq:nu_perp}) takes
the form
\begin{eqnarray}
  \label{eq:drift-parm-model}
  m_\perp=\tan\left[\arccos\left(\frac{1}{f(s)}  \right)  \right]
  \approx \sqrt{2 c s}\;,
\end{eqnarray}
where the approximations above assume $s\ll1$.

It seems likely that this simple phenomenological model applies to the
opening angles created by mutant strains of microorganisms on a Petri
dish.  Note that the square root dependence in Eq. (51) suggests that
the sector angle could be a quite sensitive measure of weak selective
differences.  Other functional relations between and s are possible,
however.  As discussed in Appendix B, when number fluctuations are
strong compared to the selective advantage at the front (strong noise
limit), a linear relation arises in \emph{two}-dimensional stepping stone
models.  These models allow some inter-diffusion of mutant and wild
type strains after the population wave has passed by.  A linear
relation can also arise for the stochastic Fisher genetic waves
generated associated with weakly deleterious mutations. 

It remains to be seen experimentally which model most accurately
describe the function $m_\perp(s)$ in a given situation.  However, our
parameterization of beneficial mutations in terms of the
phenomenological parameter $m_\perp$ is quite generally applicable to
any growth scheme leading to a bias $m_\perp$ in the random walk of
domain boundaries.  In the absence of a detailed microscopic
understanding, one can always choose to parameterize beneficial and
deleterious mutations directly in terms of $m_\perp$ itself.  Although
the arguments above focus on $m_\perp>0$, it is easy to see that
deleterious mutations are described by $m_\perp<0$.  In the following,
we analyze the evolutionary consequences of this bias for both
positive and negative values of $m_\perp$.

\subsubsection{Beneficial mutations, $m_\perp>0$}
\label{sec:benef-mutations}
Isolated beneficial mutations are often lost due to genetic drift.
However, because a sector of beneficial mutation tends to increase in
size, $m_\perp>0$, its survival probability is increased compared to
the neutral case. Here, we determine the fixation probability of a
sector of beneficial mutations, and derive how frequently those
sectors appear, given a certain rate of beneficial mutations.

Similar to the unbiased case, we will determine the fixation
probability from a diffusion equation that describes the statistical
properties of the size $X$ of a sector. Let $u(r|x_0,r_0)$ be the
probability that a sector of beneficial mutations reaches fixation up
to and including front position $r$ given that it had size
$X(r_0)=x_0$ at initial position $r_0$. Using analogous arguments to
those in Sec. \ref{sec:fixation}, it can be shown that the assumptions in
Eq.~(\ref{eq:msd-selection}) lead to a {\it biased} diffusion equation
for the fixation probability $u(r|x_0,r_0)$, which reads
\begin{eqnarray}
  \label{eq:pde-surv-selection}
  \partial_{r_0} u(r|x_0,r_0)&=&- 2 D_X\partial_{x_0}^2 u(r|x_0,r_0)
  -2m_\perp\partial_{x_0}  u(r|x_0,r_0)\;. 
\end{eqnarray}
Apart from the new drift term $\propto m_\perp$ this equation is
identical to the unbiased case Eq.~(\ref{eq:pde-surv}) with the
constant diffusion parameter appropriate to linear inoculations.  We
again impose the boundary conditions of Eqs.~(\ref{eq:annihi-bc})
accounting for annihilation at $x=0$ and fixation at $x=L$, with the
periodic boundary conditions appropriate to growth along a cylinder.

Because the diffusion constant and drift parameter are both independent of
the time-like frontier position variable in the linear inoculation, we
seek a $r_0$-independent solution of Eq.~(\ref{eq:pde-surv}).
After setting the right-hand side to zero and integrating twice, we
find the steady state solution in the limit of long times, $r(t)\to \infty$,
\begin{equation}
  \label{eq:linear-inoc-surv-prob}
  u_\pa(\infty|x_0,r_0)=\frac{1-e^{-m_\perp
      x_0/D_X}}{1-e^{-m_\perp L/D_X}} \;,  
\end{equation}
which is the ultimate survival probability of a beneficial mutation at
a linear front~\footnote{Eq.~(\ref{eq:linear-inoc-surv-prob}) is
  formally similar to Kimura's fixation probability of a beneficial
  mutation in a well-mixed population~\citep{RefWorks:113} when we
  identify the product of population size and selection coefficient
  with $m_\perp L/ 2D_X$ and the frequency of the beneficial allele
  with $x/L$.}.


The exponential dependence of Eq.~(\ref{eq:linear-inoc-surv-prob}) on
$x_0$ implies that a sector almost certainly overcomes stochastic loss
when it reaches a size larger than an ``establishment length''
$l\equiv D_X/m_\perp$.  If a sector is much smaller than this
characteristic length, $x_0\ll l$, the survival probability takes the
simple form $u_\pa\approx m_\perp x_0 /D_X$ provided that $L/l\gg 1$.
This result can be used to relate the frequency at which beneficial
mutations become established in the form of sectors to the beneficial
mutation rate $\tilde \mu_b$, which has units of an inverse time.  To
this end, we assume that all beneficial mutations confer the same
selective advantage $m_\perp$, and consider the evolutionary dynamics
during short effective ``time'' increments $\Delta r$, in which
genetic drift is stronger than selection.  According to the neutral
results of Sec.~\ref{sec:neutral}, only a number $N(\Delta r)$ of
front lineages evade stochastic loss during a time increment $\Delta
r$.  Equivalently we can say that there is at any time a set of
$N(\Delta r)$ individuals whose descendants will be present after the
next time increment $\Delta r$.  Among this population of founders,
beneficial mutations occur at rate $ \mu_b N(\Delta r)$, where $
\mu_b=\tilde \mu_b/v$ is the beneficial mutation rate in units of an
inverse length.  These mutations will \emph{ultimately} survive at
long times with probability $m_\perp x_0/D_X$, where $x_0=L/N(\Delta
r)$ is the average size of a sector after time $\Delta r$.
Multiplying these factors together yields the rate at which beneficial
mutations become established
\begin{equation}
  \label{eq:frequ-beneficials}
  \mbox{Beneficial mutations establishment rate}= \mu_b L m_\perp/D_X  \;.
\end{equation}
Note that the $\Delta r$-dependent sector number $N(\Delta r)$ drops
out of the final result~\footnote{Note that our argument assumes that
  the width of sector boundaries is smaller than the establishment
  length $l$. Otherwise, it is not possible to describe the sectoring
  dynamics as a random walk of boundaries when they are closer than
  $l$.}. Note also that the establishment rate is an inverse length in
our choice of units, which has the following interpretation: It
measures the number of beneficial mutations appearing per unit length
of progression of the front.  As discussed below, in situations where
beneficial mutations are of the type proposed in
Fig.~\ref{fig:beneficial-mutation}, Eqs.~(\ref{eq:theta}) and
(\ref{eq:frequ-beneficials}) can form the basis of an experimental
technique to measure the relative fitness advantage of beneficial
mutations (as embodied in the parameter $m_\perp$) as well as
beneficial mutation rates in microbial colonies.

\subsubsection{Deleterious mutations, $m_\perp<0$}
\label{sec:delet-mutations}
\begin{figure}
  \center
  \includegraphics[scale=0.6]{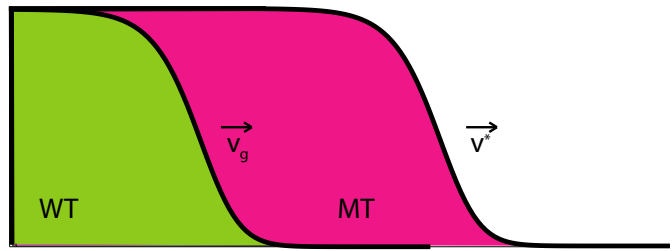}
  \caption{ A deleterious mutation can permanently take over a
    population front in one spatial dimension if the speed of the
    genetic wave $v_g$ of the wild type (WT) invading the mutants (MT)
    is lower than the speed $v^\star$ of the advancing population
    front of mutants.}
  \label{fig:delet-mutation-1d}
\end{figure}

Given the importance of chance effects during population expansions,
one may wonder to what extent deleterious mutations can prevail at
expanding frontiers. Indeed, a recent simulation
study~\citep{travis-07} has observed deleterious mutations that are
swept to high frequencies by population waves.  It is easy to see how
these ``gene surfing'' events of deleterious mutations may come about
in one spatial dimension.  The first step is a matter of chance. A
deleterious mutation needs to arise close to the wave front and become
frequent there, despite natural selection. Such surfing
\emph{attempts} are promoted by the strong genetic drift at expanding
frontiers, which are characterized by small effective population
sizes~\citep{halla-nelson-TPB-2007}.  Once the front has been taken
over by a deleterious mutation, the population wave advances into
empty territory at a somewhat reduced velocity $v^\star<v$ due to the
reduced fitness of the mutants heading the expansion.  The wild type
population, on the other hand, is stuck in the bulk of the population,
but nevertheless advances by displacing the less fit mutants. The
ensuing genetic wave of advance may be described by a Fisher genetic
wave~\citep{fisher37} with velocity $v_g$ (see Fig.
\ref{fig:delet-mutation-1d}).  Although the wild type population tries
to catch up with the front via this genetic wave, this will actually
never happen if the genetic wave is slower than the population wave of
the mutants, $v_g<v^\star$.  Thus, deleterious mutations may take over
permanently if $v^\star>v_g$. See Supplementary Text 2 for
an example of a \emph{one} dimensional model that exhibits this
behavior over a broad parameter range.

\begin{figure}
  \center
  \includegraphics[scale=0.6]{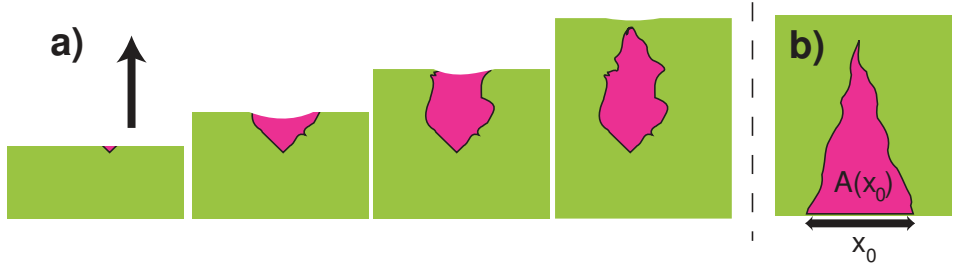}
  \caption{ a) A deleterious mutation (red) arising close to the
    population front (black lines) achieves high frequency by
    temporarily ``surfing'' at the front. Figure b) defines the
    quantity $A(x_0)$, which measures the expected area generated by a
    mutant sector if it has initial size $x_0$. }
 \label{fig:deleterious-mutation}
\end{figure}

In two spatial dimensions, isolated deleterious mutations cannot
permanently prevail at expanding frontiers.  Although a sector
harboring deleterious mutations may arise by chance effects as well,
this sector is ultimately doomed to extinction because it has a lower
expansion velocity than the surrounding wild type sectors.  As a
consequence, it is just a matter of time until the deleterious mutants
are literally overtaken by the wild type population, see
Fig.~\ref{fig:lin-inocu-scheme} and the time sequence illustrated in
Fig.~\ref{fig:deleterious-mutation}a. Nevertheless, deleterious
mutations that temporarily surf on a population wave - until they
finally ``fall off'' the wave front - pose a potentially serious
threat to the pioneer population because they achieve much higher
frequencies than expected under well-mixed conditions. As we
demonstrate below, when these deleterious alleles become numerous,
they can even trigger mutational meltdown in a cooperative manner.

We seek to quantify the frequency of surfing deleterious mutations at
expanding frontiers within the diffusion approximations used in this
paper. As before, let $F(x,r|x_0,r_0)$ be the probability that a deleterious
sector has size $X(r)=x$ at frontier position $r$, given that it had
size $X(r_0)=x_0$ at earlier ``time'' $r_0$.  The biased diffusion
equation for this distribution function $F(x,r|x_0,r_0)$ reads
\begin{equation}
  \label{eq:backw-kolmogorov}
  \partial_{r_0} F(x,r|x_0,r_0)=-2D_X\partial_{x_0}^2 F(x,r|x_0,r_0)+2\left |
    m_\perp \right | \partial_{x_0}
  F(x,r|x_0,r_0) \;.
\end{equation}
This equation generalizes the unbiased Eq.~(\ref{eq:diff-eq}),
formulated backwards in time, and has the same form as
Eq.~(\ref{eq:pde-surv-selection}) for the fixation probability
$u(r|x_0,r_0)$ of a beneficial sector. The only difference is that the
drift term has the opposite sign describing the shrinking of
deleterious sectors.

We will now use equation Eq.~(\ref{eq:backw-kolmogorov}) to study the
statistical properties of the \emph{area} $A(x_0)$ spanned by a
deleterious sector of initial size $x_0$ (see Fig.
\ref{fig:deleterious-mutation}b). When combined with the mutation
rate, this quantity turns out to control the steady state frequency of
deleterious mutations at an advancing population front.  On the
``space-time'' plot of two random walkers, where $r$ is the time-like
coordinate in the growth direction, $A(x_0)$ simply represents the
area enclosed by the two colliding world lines, given they are
initially separated by a distance $x_0$.  The average area
$\avg{A(x_0)}$ can be expressed as an integral of $F(x,r|x_0,r_0)$
over its final coordinates,
\begin{equation}
  \label{eq:area-of-del-sector}
  \avg{A(x_0)}=\mint{dr}{r_0}{\infty} \mint{dx}{0}{\infty} x F(x,r|x_0,r_0)\;.
\end{equation}
To obtain a differential equation for $\avg{A(x_0)}$, we multiply
Eq.~(\ref{eq:backw-kolmogorov}) by $x$ and integrate over $x$ and $r$,
\begin{equation}
  \label{eq:M-eqn-1}
  \mint{dr}{r_0}{\infty}  \mint{dx}{0}{\infty} x \,\partial_{r_0}
  F(x,r|x_0,r_0)=-2D_X\partial_{x_0}^2 
  \avg{A(x_0)}+2\left | m_\perp \right | \partial_{x_0}    \avg{A(x_0)} \;.
\end{equation}
Because the left hand side represents the total derivative
$\partial_{r_0} \avg{A}$ up to a boundary term evaluated at $r=r_0$,
we have
\begin{equation}
  \label{eq:M-eqn}
  \partial_{r_0} \avg{A(x_0)}+\mint{dx}{0}{\infty} x
  F(x,r_0|x_0,r_0)=-2D_X\partial_{x_0}^2 \avg{A(x_0)}+2\left | m_\perp
  \right | \partial_{x_0}  \avg{A(x_0)} \;. 
\end{equation}
Upon noting that $F(x,r_0|x_0,r_0)=\delta(x-x_0)$ and that the
expected mutant area $\avg{A}=\avg{A(x_0)}$ is $r_0$-independent in
linear inoculations, we obtain a simple ordinary differential equation
for $\avg{A(x_0)}$,
\begin{equation}
  \label{eq:M-eqn}
  2D_X\partial_{x_0}^2 \avg{A(x_0)}-2\left | m_\perp \right |
  \partial_{x_0} \avg{A(x_0)} 
  +x_0=0 \;. 
\end{equation}
This equation is solved by
\begin{equation}
  \label{eq:M-solution}
  \avg{A(x_0)}=\frac{x_0^2}{4\left |m_\perp\right |}+\frac{D_X x_0}{2m_\perp^2}\;.
\end{equation}
The quadratic contribution to the area, which dominates for $x_0\gg
l=D_X/|m_\perp|$, is just the deterministic expectation for the area
of the mutants neglecting genetic drift: With no diffusion of its
boundaries, the deleterious sector should shrink laterally at a rate
$2|m_\perp|$ and thus collapse when the front has advanced by
$r_c=x_0/(2|m_\perp|)$.  The deterministic sector area should thus
equal the area $x_0^2/4|m_\perp|$ of an isosceles triangle with height
$r_c$ and base $x_0$. Note that the characteristic length
$l=D_X/|m_\perp|$ now plays the role of a ``disestablishment
length''.

The linear part in Eq.~(\ref{eq:M-solution}) is due to stochastic
fluctuations and dominates on small scales, for $x_0\ll l$. As we now
show, this stochastic part determines the \emph{mutational load} of an
expanding population. As in Sec.~\ref{sec:benef-mutations}, we note
that stochastic fluctuations dominate over selection on a small ``time''
scale $\Delta r$. There are at any time $N(\Delta r)$ founders -
mutants or wild type - that give rise to sectors of average size
$x_0=L/N(\Delta r)$ at a small time $\Delta r$ in the future. Among
these $N(\Delta r)$ individuals deleterious mutations occur at a rate
$N(\Delta r)  \mu_d$, with $ \mu_d=\tilde \mu_d/v$ representing
the effective deleterious mutation rate per unit length of frontier
growth. This process produces deleterious mutants at a rate given by
\begin{equation}
  \label{eq:gamma}
    \mu_d N(\Delta r) \avg{A[L/N(\Delta r)]}\equiv \gamma L
\end{equation}
For $\Delta r\to 0$, the short time neutrality assumption becomes
exact, the factors of $N(\Delta r)$ cancel and we obtain
\begin{equation}
  \label{eq:gamma-2}
  \gamma= \frac{D_X \mu_d}{2 m_\perp^2} \;.
\end{equation}
The restrictions of the footnote after
Eq.~(\ref{eq:frequ-beneficials}) apply here as well, and we have
assumed for simplicity that all deleterious mutations confer the
same selective disadvantage $-|m_\perp|$.

The parameter $\gamma$ characterizes the mutational load. If
$\gamma\ll1 $, it represents the fraction of the individuals in the
newly colonized regions that carry the deleterious mutation. When the
parameter $\gamma$ becomes of order one, mutant sectors become so
numerous that they start to collide. These collisions are not captured
by our simple theory, so $\gamma$ can no longer be interpreted
literally as the fraction of mutants for $\gamma=O(1)$.  Still,
$\gamma$ defined by Eq.~(\ref{eq:gamma-2}) can be viewed as a
dimensionless control parameter that determines whether the pioneer
population remains close to wild type or the mutational load becomes
so strong that the average fitness deteriorates dramatically.

\begin{figure}
  \psfrag{mperpmperp=0.1}{$m_\perp=0.1$}
  \psfrag{mperpmperp=0.01}{$m_\perp=0.01$}
  \psfrag{mperpmperp=0.04}{$m_\perp=0.04$}
  \psfrag{gamma}{$\gamma= \mu_d D_X/(2m_\perp^2)$}
  \psfrag{pwt}{$p_{WT}$}
  \center
  \includegraphics[scale=0.5]{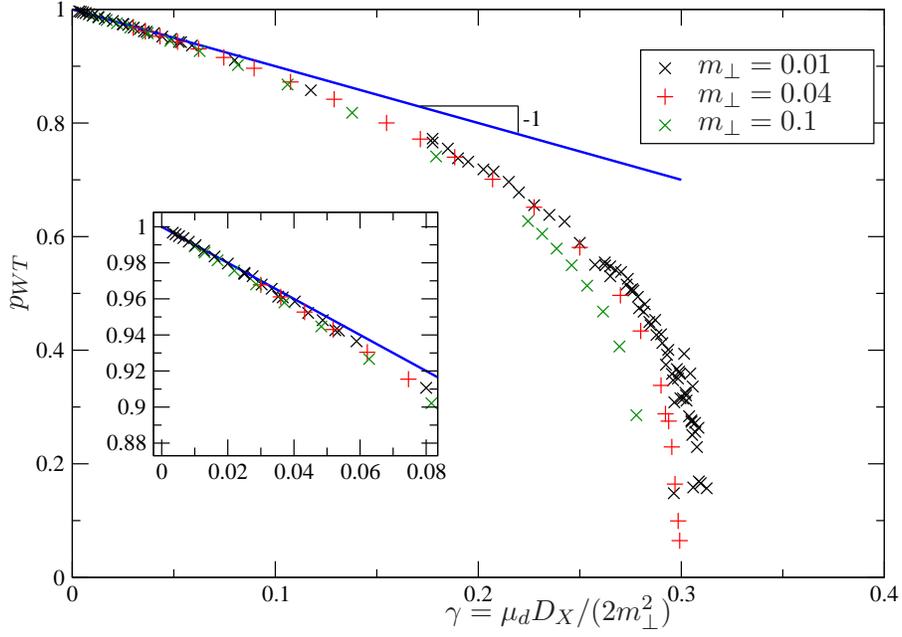}
  \caption{ Simulation results for the fraction $p_{WT}$ of the wild
    type population for various values of the drift parameter
    $|m_\perp|$, characterizing the selective advantage of the wild type
    population, as a function of the dimensionless control parameter
    $\gamma= \mu_d D_X/(2m_\perp^2)$. The simulations were
    carried out for small values of $m_\perp$ because our theoretical
    treatment applies to the limit of weak selection.  To limit
    computational intensity, we restricted the simulations to
    $m_\perp>0.01$.  Upper figure: With decreasing $m_\perp$,
    $p_{WT}(\gamma)$ seems to approach a universal function. The
    simulations confirm our analytical result that $1-p_{WT}\sim
    \gamma$ as $\gamma\to 0$, see inset. On the other hand, they also
    show that $p_{WT}\to 0$ above a critical threshold. The lowest
    probed value of $m_\perp=0.01$ indicates $\gamma_c =0.32\pm 0.02$.
    The simulation algorithm is described in
    Supplementary Text 3.}
  \label{fig:error-threshold}
\end{figure}

In fact, simulations for the simple model described in
Supplementary Text 3 do indeed reveal a critical value
$\gamma_c=0.32\pm0.02$ that separates two regimes, see
Fig.~\ref{fig:error-threshold}.  For $\gamma<\gamma_c$, the average
fraction of wild type surviving in a pioneer population stabilizes at
a finite steady state value, whereas for $\gamma>\gamma_c$ the pioneer
population keeps accumulating deleterious mutations until the wild
type is lost entirely.  This apparent continuous phase transition is
the one dimensional spatial analog of the error threshold in
well-mixed populations~\citep{EIGEN::58:p465:1971}.  Note that,
depending on the functional relation between $m_\perp$ and the
selective disadvantage $s<0$, the threshold value $\gamma_c$ for
expanding populations can be very different from the well-known
prediction $\gamma_c^{w.m.}=\tilde \mu_d/|s|=1$ for well-mixed populations.
For the ``geometric model'' outlined at the beginning of
Sec.~\ref{sec:nat-selection}, we have $m_\perp\propto \sqrt{s}$ and
thus $\gamma_c\propto s^{-1}$ similar to a well-mixed population.
However, for a two-dimensional stepping stone model in the strong
noise limit, the relationship between $m_\perp$ and selective
disadvantage $s$ is linear (cf. Supplementary Text 2).
In this case, the genetic load $\gamma\propto s^{-2}$ could be
strongly increased due to weakly deleterious alleles.

A simple argument suggests that the collision of mutant domains will
indeed increase the expected fraction of mutants.  Consider the
collision of two mutant domains of sizes $x_1$ and $x_2$,
characterized by the same selective disadvantage $|m_\perp|$. Then the
newly combined domain has the increased size $x_1+x_2$.  The expected
area of the combined region as given by Eq.~(\ref{eq:M-solution}) is
\emph{larger} than the two separated domains by an amount
\begin{equation}
  \avg{A_{ex}(x_1,x_2)}=\avg{A(x_1+x_2)}-\avg{A(x_1)}-\avg{A(x_2)}=\frac{x_1 
    x_2}{2|m_\perp|}   \label{eq:excess-area} \;. 
\end{equation}
This excess mutant area generated by the collision of domains
furnishes a cooperative \emph{interaction} associated with the mutational
meltdown, as revealed in Fig.~\ref{fig:error-threshold} as
$\gamma\to\gamma_c$.

\section{Conclusions and discussion}
\label{sec:discussion}

%

We have studied the impact of a continuous range expansion on the
evolutionary dynamics of spatially structured populations. Our
analysis focuses on the enhanced genetic drift at advancing frontiers,
which leads to strong genetic differentiation. Most directly, our
results apply to recent experiments on expanding microbial
colonies~\citep{OskarHallatschek12042007}. However, we believe that the
present analysis is of more general significance for populations that
grow continuously and isotropically in two dimensions. The advancing
frontier of those populations also has a thin layer of active
pioneers, whose width depends on demographic parameters of the range
expansion, such as the growth, migration and turnover rates.  As the
population advances, the pioneer population in this quasi-one
dimensional front region is continuously re-sampled, as in a
one-dimensional stepping stone model~\citep{KIMURA:W::49:p561:1964},
for which it is known that allelic segregation occurs locally beyond a
crossover time~\footnote{In a continuous one-dimensional stepping
  stone model, well-defined boundaries appear after the characteristic
  time $T=c_{1d}^2 D \tau_g^2$, where $D$ is the usual diffusivity of
  the individuals (with dimensions cm$^2/$sec), $\tau_g$ is the
  generation time characterizing the strength of population turnover,
  and $c_{1d}$ is the one-dimensional population density with units of
  an inverse length~\citep{Barton:D:E::61:p31-48:2002,halla-kirill-08}.
  Considering the advancing edge of the colony as an effectively
  one-dimensional habitat of width $v/a$ depending on the expansion
  velocity $v$ and effective growth rate $a$ leads to the estimate
  $c_{1d}\approx c_{2d} v/a $.  Thus, we expect segregation to occur
  on times larger than $T\approx D c_{2d}^2 v^2 \tau_g^2/a^2$.
  Domains will then occur on length scales larger than $w=\sqrt{D
    T}\approx D c_{2d}v \tau_g/a$, where $w$ represents a
  characteristic width of domain boundaries.}. Therefore, we believe
that the coarsening process, and our model for it in terms of sharp
domain boundaries, might be generally important for homogeneously
expanding populations.

Within this model for population dynamics and genetic segregation of a
continuous two-dimensional range expansion out of some prescribed
initial habitat (linear or circular), we first studied the
\emph{neutral} genealogies of single loci without mutations. We found
that only a fraction of founder cells are able to propagate their
genes with the advancing front. Due to population turnover in the thin
band of pioneers, the number of lineages gradually decreases as the
expansion progresses. If a linear front advances by a distance $\Delta
r$, the number of survivors decays like $1/\sqrt{\Delta r}$. One
sector will dominate after an average fixation time $\propto L^2$,
where $L$ is the linear dimension of the front.  In the case of
circular colonies, we find instead that a finite number of sectors
survives, due to the geometric expansion of the perimeter which
opposes genetic drift.  For our moel, the expected number of these
surviving sectors is proportional to the inverse square root of the
initial radius of the colony.  Both results assume that domain
boundaries carry out a diffusive random walk.  Growth models with
surface roughness are expected to show an anomalously vigorous wall
wandering~\citep{RefWorks:49,OskarHallatschek12042007} where the
variance in separation $X(r)$ between two walls grows as $(\Delta
r)^{2\zeta}$ with an exponent $2 \zeta>1$ as the front advances by a
distance $\Delta r$.  Our analysis of annihilating random walkers
could be extended to these cases by assuming a diffusion ``constant''
that decreasing with the separation between the two random
walkers~\footnote{At a linear front, for instance, the diffusion
  constant $D_X(\Delta r)\sim (\Delta r)^{2-\zeta^{-1}}$ describes a
  super-diffusive random walk with exponent $\zeta$.}.  On the scaling
level, such an analysis would yield an average sector number for
linear inoculations increasing like $(\Delta r)^{-\zeta}$ instead of
$(\Delta r)^{-1/2}$. For radial inoculation, the number of surviving
sectors $N(\infty, r_0)$ (see Eq.~(\ref{eq:n-approx})) now scales with
the initial radius according to $N(\infty, r_0)\sim r_0^{1-\zeta}$. We
speculate that the effect of inflation balancing genetic drift could
be relevant for explaining the surprisingly large levels of genetic
diversity among some invasive species, that have been introduced
locally into a new and favorable habitats (e.g.  rabbits in
Australia). According to our model for radial expansions, those
species are expected to preserve a considerable amount of their
initial diversity during the habitat expansion.

Next, we considered the evolutionary dynamics of beneficial mutations
arising at frontiers.  To this end, we modified our model by adding a
deterministic bias to the sectoring dynamics that tends to increase
the size of sectors.  This modification applies most directly to
beneficial mutations whose only effect is to increase the growth rate
of individuals $v\to v^\star=f(s) v $.  We found that, at a linear
frontier, a successful beneficial mutation ``emits'' a sector that has
an opening angle $\Phi$ given by Eq.~(\ref{eq:theta}). If the change
in velocity due to the mutations is weak, $s\ll 1$, then the sector
angle is approximately given by $\Phi\approx 2 \sqrt{2 c s}$, where
$c=1/2$ or $c=1$ depending on wether genetic drift is weak or strong
at the front, respectively.  The square root dependence on the
selection coefficient indicates that sector angles are a sensitive
measure of fitness differences. Thus, measuring sector-angles could be
a useful tool to decipher the distribution of fitness effects of
beneficial mutations, which can be difficult (or tedious) to measure
by liquid culture techniques.  The use of advancing population waves
in evolutionary studies has been demonstrated earlier in a
one-dimensional study of the effect of beneficial mutations on
replicating RNA molecules~\citep{mccaskill1993ieo}. In this study,
beneficial mutations were identified by a spontaneous increase in
expansion velocity of a Fisher population wave of proliferating RNA
molecules travelling along a one-dimensional tube. In the
two-dimensional extension of these studies, we expect a marked
increase in resolution because we expect sector angles to be much more
sensitive to selective changes than spreading velocities.

Although our parameterization of beneficial mutations in terms of the
phenomenological parameter $m_\perp$ is quite generally applicable,
the relation between bias $m_\perp$ and fitness difference remains to
be assessed experimentally.  The example of a beneficial mutation in
Fig.~\ref{fig:micrographs-beneficial-mutation} indeed seems to
approach an asymptotic sector angle. The sector shape deviates
initially somewhat from the simple triangular sector geometry used in
our purely geometric model, however. This transient feature suggests
that a number of other factors could change the relation between bias
$m_\perp$ for small sectors and the fitness difference in microbial
populations~\citep{RefWorks:45}, such as a line tension penalizing
front deformations. A related possibility is an effective attraction
between oppositely directed Fisher genetic waves that slows down their
separation at early times.

Finally, we estimated the mutational load in an expanding population.
Deleterious mutations were found to proliferate at expanding frontiers
rather differently than in well-mixed populations. Given that
deleterious mutations with a selective disadvantage $|m_\perp|$ occur
at a small rate $\tilde \mu_d$, we determined the fraction $\gamma$ of
mutants at a linear wave front as $\gamma\equiv D_X \tilde \mu_d/(2 v
m_\perp^2)$.  Here, $v$ is the expansion velocity of the population
wave, $D_X$ is the wall diffusion constant (in units of
length$^2/$length) and $m_\perp$ represents the ``speed'' or slope by
which a single deleterious mutation is squeezed out of the population
front\footnote{$|m_\perp|$ is approximately given by half the
  ``closing'' angle of a sector harboring deleterious mutations.}.
Furthermore, we presented numerical evidence that populations suffer
from genetic meltdown as $\gamma\to\gamma_c\approx 0.32\pm 0.02$. This
\emph{error threshold} should be compared with the expectation
$\gamma_{w.m.}=\tilde \mu_d/s$ for well-mixed
populations~\citep{EIGEN::58:p465:1971}, where $s$ is the relative
fitness detriment of the deleterious mutations. Depending on the
product $\gamma/\gamma_{w.m.}=D_X s/(2 v m_\perp^2)$, both predictions
can be very different. If wall wandering represented by $D_X$ is
strong, or the bias $m_\perp$ for a given fitness difference is weak
(for instance $m_\perp\propto s$ as in the weak selection limit of the
stepping stone model, cf.~Supplementary Text 2, then the
genetic load during a range expansion becomes substantially larger
than in the well-mixed case. Thus, deleterious mutations could
accumulate quite strongly during times of habitat expansions, thereby
setting tight constraints on the dynamics of range expansions. Genetic
load could be serious threat during general species invasions. This
effect might extend to past range expansions of humans, as the
mutational load inside Africa was found to be significantly lower than
in the more recently colonized Europe~\citep{RefWorks:104}.

{\bf Acknowledgments:} It is a pleasure to acknowledge helpful
conversations with Nilay Karahan, Andrew Murray, Sharad Ramanathan,
John Wakeley, and, in particular, Kirill Korolev, who suggested
important corrections to the manuscript. The haploid strains (mating
type \emph{a}) of \emph{S.  cerevisiae} used for
Fig.~\ref{fig:micrographs-beneficial-mutation} were obtained through
the generosity of John Koschwanez (FAS Center for Systems Biology,
Harvard University), and are derived from a W303 strain and have
either Cerulean (CFP) or mCherry (RFP) constitutively expressed from
the ACT1 promoter and integrated at the ACT1 promoter locus. This
research was supported by the German Research Foundation through grant
no.~Ha 5163/1 (O.~H.), the National Science Foundation through Grant
DMR-0654191, a National Institute of General Medical Sciences Grant,
and the Harvard Materials Research Science and Engineering Center
through Grant DMR-0213505 (D.~R.~N. and O.~H.). Simulations were
performed at the Center for Nanoscale Systems (CNS), a member of the
National Nanotechnology Infrastructure Network (NNIN), which is
supported by the National Science Foundation under NSF award no.
ECS-0335765.

\appendix

\section{Diffusion equation for periodic boundary conditions}
\label{sec:exact-solution}
In the main text, we gave a simple argument how to solve the diffusion
equation (\ref{eq:diff-eq}) for annihilating random walks with
absorbing boundary conditions at $z=0$. However, this solution is only
strictly valid in unbounded space. Here, we derive the slightly more
complicated solution Eq.~(\ref{eq:phi-exact}) valid for finite
systems.  This solution satisfies another absorbing boundary
condition, Eq.~(\ref{eq:boundary-condition-L}), which accounts for the
fixation when the sector reaches the size of the system $L$.

We make the following ansatz
\begin{eqnarray}
  \label{eq:F-ansatz}
  F(z,r|z_0,r_0)&=&\sum_{n=1}^{\infty}a_n(r)W_n(z) \\ 
  a_n(r)&=&\mint{dz}{0}{ L}W_n(z) F(z,r|z_0,r_0)\;,
\end{eqnarray}
where the sine mode $W_n(z)$ and wave numbers $q_n$ were defined in
Eq.~(\ref{eq:sine-modes}).  This expansion in terms of sine-modes on
the right-hand-side guarantees both boundary conditions.

Observe that the ansatz Eq.~(\ref{eq:F-ansatz}) solves
Eq.~(\ref{eq:diff-eq}) provided that the mode amplitudes $a_n(r)$ obey
\begin{equation}
  \label{eq:EOM-modeamplitudes}
  \partial_t a_n(r)=-2q_n^2 D(r)a_n(r) \;.
\end{equation}
Integrating this from $r_0$ to $r$ gives
\begin{equation}
  \label{eq:time-evol-mode-amplitudes}
  a_n(r)=a_n(r_0)e^{-q_n^2 \sigma^2/2}
\end{equation}
where $\sigma$ is the standard deviation in the $z$ coordinate, as
defined in Eq.~(\ref{eq:stddev-general}). The pre-factor is determined
from the initial condition $F(z,r_0|z_0,r_0)=\delta(z-z_0)$,
\begin{eqnarray}
  \label{eq:final-cond}
  a_n(r)&=&\mint{dz}{0}{L}W_n(z)F(z,r|z_0,r_0) \\
  &=&\sqrt{\frac 2L} \sin(q_n z_0)\;.
\end{eqnarray}
Hence, the solution for the probability distribution reads
\begin{equation}
  \label{eq:sect-size-distrib}
  F(z,r|z_0,r_0)=\frac 2L \sum_{n=1}^{\infty}\sin(q_n z)\sin(q_n z_0)
  e^{-q_n^2 \sigma^2/2}\;,
\end{equation}
which is Eq.~(\ref{eq:phi-exact}). 

Next, we would like normalize the distribution $F(z,r|z_0,r_0)$ by the
total probability $\bar u(r|z_0,r_0)$ that a sector of initial size
$z_0$ has not collapsed. This probability can be written as a sum of
two terms,
\begin{eqnarray}
  \label{eq:normali-cond-a}
  \bar u(r|z_0,r_0)&=&u(r|z_0,r_0)+\mint{dz}{0}{L} F(z,r|z_0,r_0) \\
  &=&\frac{z_0}L+\frac2L \sum_{n=1}^{\infty}\frac{\sin(q_n z_0)}{q_n}e^{-q_n^2
    \sigma^2/2}\;.
    \label{eq:normali-cond-b}
\end{eqnarray}
The first term on the right hand side in Eq.~(\ref{eq:normali-cond-a})
is the probability that the sector has reached fixation, which was
evaluated in Eq.~(\ref{eq:surv-proba}), and the second term is the
total probability that a sector has neither collapsed nor reached
fixation.  In the limit $z_0\to 0$, we can now construct the
normalized distribution
\begin{equation}
  \label{eq:sect-size-dist-normalized}
  P(z,r|r_0)=\lim_{z_0\to0}
  \frac{F(z,r|z_0,r_0)}{\bar u(r|z_0,r_0)}=
  \frac{\sum_{n=1}^{\infty}2q_n \sin(q_n z)   e^{-q_n^2 \sigma^2/2}}
  {\vartheta_3\left[0,\exp\left(\sigma^2 \pi^2/(2 L^2)\right)\right]} \;.
\end{equation}
This limiting distribution is depicted in Fig.~\ref{fig:sprob} for
various values of $L/\sigma$. Note that the results are
indistinguishable from the asymptotic result
Eq.~(\ref{eq:sector-size-distribution}) for $L/\sigma\leq 4$.  The
area under these plots represent by construction the probability that
a randomly chosen sector has not yet reached fixation.

\section{Genetic versus population waves}
\label{sec:surfing-delet-one-dim}

In this Appendix, we first exhibit an explicit model for the surfing of a
deleterious gene in one dimension. Unlike the two-dimensional case
considered in Ref.~\cite{travis-07}, we focus here on one dimension,
leading to the situation sketched Fig.~\ref{fig:delet-mutation-1d}.
Provided we focus on populations that stop growing and diffusing once
the population has passed by when $c+c^\star=1$ (see below), we can view
this model as an approximation to the dynamics within the thin layer
of actively growing pioneers in a two-dimensional range expansion, as
in Fig.\ref{fig:micrographs-neutral}a) and
\ref{fig:micrographs-neutral}d). This approximation neglects number
fluctuations at the front. The time variable $t$ in this Appendix then
corresponds to the frontier position $r$ used elsewhere in this paper.

\begin{figure}[tbh]
  \center
  \includegraphics*[scale=0.5,clip=true]{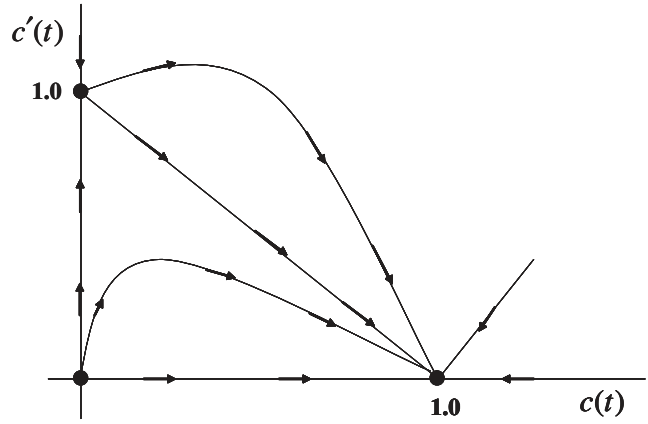}
  \caption{Dynamics associated with Eqs.~(\ref{eq:fixpoints}) in zero
    dimensions, e.g., in a well-mixed overnight culture of microbes
    where spatial diffusion plays no role.}
  \label{fig:flow-diagram}
\end{figure}

Let $c(x,t)$ be the dimensionless concentration of wild type
individuals (growth rate $a$), and $c^\star(x,t)$ the dimensionless
concentration of the mutant strain (growth rate $a^\star$).  We assume
for now $0<a^\star<a$, so that the mutation is deleterious relative to
the wild type. We work in the strong selection limit, and neglect
fluctuations in the number of discrete individuals within the two
populations, although these can be important under some circumstances.
For simplicity, we assume identical diffusion constants and a common
steady state value or ``carrying capacity'' of unity in rescaled units
for these two populations, both separately and when they are mixed
together. The two populations differ in their growth rates, and in
addition the wild type secrets a chemical that impedes the growth of
the mutant under crowded conditions. A simple set of coupled
reaction-diffusion equations for the two strains then reads,
\begin{eqnarray}
  \label{eq:coupled-fisher}
  \partial_t c&=&D\partial_x^2 c + a c (1-c-c^\star)-\alpha  c^\star c
  \;,\\
  \partial_t c^\star&=&D\partial_x^2 c^\star + a^\star c^\star
  (1-c^\star-c)+ \alpha c c^\star \;.\nonumber
\end{eqnarray}
``Crowded conditions'' corresponds to $c+c^\star=1$, and the term
$-\alpha c^\star c$ (with $\alpha>0$) in the first equation represents
the secretion of an inhibitory chemical. Consider first the
zero-dimensional case of well-mixed, spatially uniform populations, so
that $c$ and $c^\star$ are a function of time only.  It is
straightforward to show that evolution of $c(t)$ and $c^\star(t)$ is
then controlled by three fixed points, namely
\begin{eqnarray}
  \label{eq:fixpoints}
  (0,0)\mbox{, eigenvalues } a \;\&\;  a^\star \mbox{ (unstable)}\nonumber\\
  (0,1)\mbox{, eigenvalues } -a\; \&\; \alpha \mbox{ (hyperbolic)}\\
  (1,0)\mbox{, eigenvalues} -a^\star\; \&\; -\alpha \mbox{ (stable)}\nonumber
\end{eqnarray}
and that the invariant subspaces are $c = 0, c^\star= 0, c + c^\star =
1$. The equal and opposite coupling strengths of the last terms in
Eqs.~(\ref{eq:coupled-fisher}) were chosen to insure that $c + c^\star
= 1$ is an invariant subspace, and that we deal with the relatively
simple situation that a composite population has the same carrying
capacity as either population in isolation. The dynamics associated
with this well-mixed effectively zero-dimensional population is shown
in Fig.~\ref{fig:flow-diagram}.  If $a^\star\approx a > 0$, both
populations will grow up rapidly after a small inoculation near
$(0,0)$. However, provided $0 < \alpha \ll 1$ , the ultimate fate of
the population is then a slow drift down the line $c + c^\star\approx
1$ until the wild type dominates at the stable fixed point $(1,0)$.

Now consider what happens in one spatial dimension. If the mutant
strain is absent, ($c^\star(x,t)=0$), standard
considerations~\cite{murray-book-chap11-fisherwave} show that the remaining equation,
\begin{equation}
  \label{eq:fisher-eqn}
  \partial_t c = D\partial_x^2 c+ a c(1-c)
\end{equation}
supports stable left- and right-moving waves that interpolate between
$c(x,t)=0$ and $c(x,t)=1$ of the form
\begin{equation}
  \label{eq:wave-eqn}
  c(x,t)=f(x\pm vt) \;,
\end{equation}
with the velocity 
\begin{equation}
  \label{eq:fisher-velocity}
  v=2\sqrt{D a}
\end{equation}
and width $w=\sqrt{D/a}$. On the other hand, if $c(x,t)=0$ everywhere,
the Fisher wave describing the mutant population $c^\star(x,t)$ has
velocity
\begin{equation}
  \label{eq:mutant-velocity}
  v^\star=2\sqrt{D a^\star}\;.
\end{equation}
Not surprisingly, the mutant Fisher population wave spreads with a lower velocity
than the wild type, $v^\star<v$.  However, consider now the situation when
\emph{both} strains are present and together always saturate the
carrying capacity of the environment, i.e., we operate along the line
$c(x,t)+c^\star(x,t)=1$. Equations Eq.~(\ref{eq:coupled-fisher}) then
collapse to a single equation. If we focus on the dynamics of the wild
type density $c(x,t)$, we now have
\begin{equation}
  \label{eq:wt-genetic-wave}
  \partial_t c=D \partial_x^2 c+ \alpha c (1-c) \;.
\end{equation}
Equation Eq.~(\ref{eq:wt-genetic-wave}) describes a Fisher genetic
wave of the wild type displacing a saturated population of mutants
with a velocity
\begin{equation}
  \label{eq:velocity-genetic-wave}
  v_g=2\sqrt{D \alpha}\;,
\end{equation}
determined by $\alpha$, which represents the selective advantage under
crowded conditions. Note that the Fisher genetic wave velocity is
proportional to the square root of this selective advantage $\alpha$.

Figure \ref{fig:delet-mutation-1d} depicts in one dimension a
superposition of a Fisher \emph{population} wave on the right, with a
mutant population moving into empty space with velocity $v^\star$, and a
Fisher \emph{genetic} wave on left, with the wild type displacing the
mutant with velocity $v_g$.  Successful ``surfing'' of a deleterious mutant
ahead of the wild type requires only that the genetic wave not
overtake the population wave, i.e., $v_g<v^\star$, which leads to the
condition
\begin{equation}
  \label{eq:condition-for-surfing-of-deleterious-mutation}
  \alpha<a^\star<a
\end{equation}
for this simplified model.

As discussed in the main text, the situation is considerably more
complicated (and interesting!) in two dimensions, and when the genetic
drift embodied in particle number fluctuations are taken into account.
Let us focus now on a \emph{beneficial} mutation forming a sector like
that in Fig.~\ref{fig:beneficial-mutation}.  We consider the simple
situation such that $a^\star\approx a$ (comparable Fisher population
wave velocities), but assume that the wildtype is favored under
crowded conditions, $\alpha<0$ with $|\alpha|\gg |a-a^\star|$. When
the above deterministic two-component system reaction-diffusion model
is generalized to two dimensions, it leads to a Fisher genetic wave
velocity $v_g=2\sqrt{D |\alpha|}$, and a Fisher population wave
velocities $v=2\sqrt{D a}\approx v^\star=2\sqrt{D a^\star}$. Assuming
these waves are configured approximately at right angles as in
Fig.~\ref{fig:beneficial-mutation}, we obtain a
$\sqrt{\alpha}$-dependence for the bias
$m_\perp=v_g/v=\sqrt{\alpha/a}$ in the diffusion of the sector size,
similar to Eq.~(\ref{eq:drift-parm-model}). In analogy with
Eq.~(\ref{eq:drift-parm-model}) in the main text (with $c = 1/2$), we
set $\alpha \equiv \tilde s a$, thus defining a selective advantage
$\tilde s$ under crowded conditions such that $a = a^\star$.

In contrast to Eq.~(\ref{eq:drift-parm-model}), a linear dependence
$m_\perp\propto \tilde s$ can result if the competition between the
mutant and wild type is weak, $\tilde s \ll 1$. In this limit, selection
is weak compared to genetic drift~\cite{RefWorks:30}, and the wave
speed of a genetic Fisher wave is linear in $\tilde s$,
\begin{equation}
  \label{eq:fisher-wave-speed-strong-noise}
  v_g=2  \tilde s a D \tau_g c_{1d} / \Delta\;, \qquad \mbox {(weak selection)} 
\end{equation}
where $D$ is the usual spatial diffusivity of the organisms, $c_{1d}$
is the effective one-dimensional population density at the frontier
and $\tau_g$ is the generation time. The parameter $\Delta$ is the
variance in offspring number in a single generation, and is in all
breeding models a number of order one.  Note that which relation
between $m_\perp$ and the selection coefficients ($\tilde s$ or $s$) is
realized strongly influences the genetic load, as predicted by
Eq.~(\ref{eq:gamma-2}).

Fisher population waves and Fisher genetic waves are approximately at
right angles in the bacterial and yeast populations of
Fig.~\ref{fig:micrographs-neutral}.  Here, $v\approx v^\star$ (so the
population fronts advance into virgin territory at a common velocity)
because these strains were chosen to be genetically neutral. There is
no competition under crowded conditions, $\tilde s=0$, and the Fisher
\emph{genetic} waves in this Figure are stalled out on average. Note
that the model Eq.~(\ref{eq:coupled-fisher}) and its generalization to
two dimensions do not apply to regions far behind the population
front of the microbiological experiments of Fig. 1., because the used
micro-organisms not only stop growing, but also stop diffusing once
the population wave has passed.

\section{Simulations of genetic load at expanding frontiers} 
\label{sec:simulations}
In this section, we describe the simulations that were used to map out
the genetic load at expanding frontiers as a function of the mutation
rate, which is reported in Fig.~\ref{fig:error-threshold}. Our
computer model is a derivative of the so-called contact process, which
is believed to belong to the universality class of directed
percolation~\cite{odor2004ucn}.

The simulation uses random sequential updates to evolve a
one--dimensional lattice of binary variables ${\sigma_i}$, $i=1\dots
N$, which can have the values WT (wild type) and MT (mutant).
Initially, all sites are WT. At each time step, a pair of sites
$\{\sigma_i,\sigma_{i+1}\}$ is chosen such that periodic boundary
conditions are respected. This pair of sites is then updated according
to $\{\sigma_{i},\sigma_{i+1}\}\to\{\sigma'_{i},\sigma'_{i+1}\}$ with
certain transition rates
$w(\sigma'_{i},\sigma'_{i+1}|\sigma_{i},\sigma_{i+1})$. To describe
the random occurrence of deleterious mutations, we implement
\begin{equation}
  \label{eq:transitionrates-drift}
  w(MT,.|WT,.)=w(.,MT|.,WT)=2D_X \tilde \mu_d \;,
\end{equation}
where the states not shown are arbitrary. To account for the biased
random walk of domain boundaries,
\begin{eqnarray}
  w(WT,WT|WT,MT)&=&w(WT,WT|MT,WT)=(1+m_\perp)/2
  \label{eq:expansion-bias}   \\
  w(MT,MT|WT,MT)&=&w(MT,MT|MT,WT)=(1-m_\perp)/2   \label{eq:shrinkage}\;.
\end{eqnarray}
All other rates are zero. The expansion of domain boundaries is
described by the first equation Eq.~(\ref{eq:expansion-bias}), and the
shrinkage by the second Eq.~(\ref{eq:shrinkage}). One lattice site in
the simulation represents a length $2D_X$ of the linear frontier, and
one time step corresponds to an equal change $\Delta r=2D_X$ in the
frontier position, or effective time. The computer model then has the
two non-dimensional parameters, $2D_X \tilde \mu_d$ and $ m_\perp$.

We explore this model in the weak selection regime where the drift
parameter $m_\perp\ll 1$ is small.  This is the regime where the
computer model generates domain boundaries that are well approximated
by slightly biased continuous time random walks, as assumed in the
main part of the paper. We find that the error threshold, at which the
wild type is lost, only depends on the parameter $\gamma=D_X \tilde
\mu_d/(2m_\perp^2)$ introduced in Eq.~(\ref{eq:gamma-2}) rather than
on the two independent parameter of the computer model separately.
This is documented by the data collapse in
Fig.~\ref{fig:error-threshold}.

\bibliographystyle{plainnat}

\bibliography{bibis/hallatschek,bibis/glacial-cycles,bibis/glacial-cycles2,bibis/bottleneck,bibis/strong-miglimit,bibis/surfing,bibis/kingman,bibis/neutral1,bibis/neutral2,bibis/selection,bibis/wright,bibis/isolation-by-distance,bibis/coyne,bibis/harris-hastings-metapop,bibis/voter,bibis/fkpp.bib,bibis/hallatschek.bib,bibis/fkpp2.bib}

\end{document}